%% file: MG_review_rs1.tex
\documentclass[preprint,aps,prd,showpacs,nofootinbib,amsmath,amssymb,amsfonts,superscriptaddress]{revtex4-1}
\usepackage{amssymb}
\usepackage{bm}
\usepackage{latexsym}
\usepackage{xcolor}
\usepackage{amsfonts,amssymb}
\usepackage{graphicx,epsfig}
\usepackage{subcaption}
\usepackage{psfrag}
\usepackage{amsmath,amssymb}
\usepackage{natbib}
\usepackage{mathrsfs}
\usepackage{amsthm}
\usepackage{lipsum}
\usepackage{array}
\usepackage{tabu}
\usepackage{multirow}
\usepackage{latexsym}
\usepackage{mathrsfs}
\usepackage{hyperref}
\usepackage{float}
\usepackage[section]{placeins}
\hypersetup{
    colorlinks=true,       
    linkcolor=red,          
    citecolor=blue,        
}

\newcommand{\addlabel}[1]{%
Eq.\refstepcounter{equation}%
\addlabel{#1}%
\let\]\endequation }
\definecolor{dgreen}{rgb}{0.00,0.60,0.00}
\newcommand*{\red}{\textcolor{red}}

\usepackage{rotating}
\usepackage{array}
\usepackage{float}
\def\a  {\alpha}
\def\b  {\beta}

\def\d  {\delta}

\def\g  {\gamma}

\def\MPl{M_{\rm Pl}}

\newcolumntype{M}[1]{>{\centering\arraybackslash}m{#1}}
\newcolumntype{N}{@{}m{0pt}@{}}

\begin{document}

\title{Modified theories of Gravity:
Why, How and What?}
%
\author{S. Shankaranarayanan}
\email{shanki@phy.iitb.ac.in}
\affiliation{Department of Physics, Indian Institute of Technology Bombay, Mumbai 400076, India}
\author{Joseph P Johnson }
\email{josephpj@iitb.ac.in}
\thanks{Corresponding author}
\affiliation{Department of Physics, Indian Institute of Technology Bombay, Mumbai 400076, India} 

\begin{abstract}
General Relativity (GR) was proven via the direct detection of gravitational waves from the mergers of the binary 
{black holes} and binary neutron stars by the Advanced LIGO and Advanced Virgo detectors. These detections confirmed the prediction of GR and provided the first direct evidence of the existence of stellar-mass black holes (BHs). However, the occurrence of singularities at the centers of BHs suggests that GR is inapplicable because of the breakdown of the equivalence principle at the singularities. The fact that these singularities exist indicates that GR cannot be a universal theory of space-time. In the low-energy limit, the theoretical and observational challenges faced by the $\Lambda$CDM model also indicate that we might have to look beyond GR as the underlying theory of gravity.
Unlike GR, whose field equations contain only up to second-order derivatives, the modified theories with higher derivative Ricci/Riemann tensor gravity models include higher derivatives. Therefore, one expects significant differences between GR and modified theories. Since there are many ways of modifying GR in the strong-gravity and cosmological distances, each model has unique features. This leads to the following crucial question: Are there a set of unique signatures that distinguish GR from modified gravity {(MG)} theories? This review discusses three aspects  of {MG} theories: 
(1) Why do we need to consider {MG theories}? (2) How to modify {GR}? and (3) What are the observational consequences? 
The review is written in a pedagogical style with the expectation that it will serve as a useful reference for theorists and observers and those interested in bridging the divide between theory and observations.
\end{abstract}

\maketitle
\section{Introduction}\label{sec1}

General Relativity (GR) has established itself as an extraordinarily successful model of gravity and cosmology. It is in remarkable agreement with a wealth of Solar System gravity precision tests, such as gravitational redshift, gravitational lensing of light from distant background stars, anomalous perihelion precession of Mercury, Shapiro time-delay effect, and Lunar laser experiments. Moreover, the above predictions are in the regime of weak gravitational fields (like near the surface of the Sun)~\cite{2006-Will-LivingRev.Rel.}. 

Outside the Solar system, predictions of GR involving changes in the orbit of binary pulsars due to gravitational wave emission and black hole (BH) mergers are being verified~\cite{2013-Stairs-LRR,2016-Abbott.etal-PRL,2017-Abbott.etal-APk}. However, the merger of two BHs is a cataclysmic event that involves strong gravity~\cite{2017-Abbott.etal-APk}. It is unclear whether GR can accurately describe gravity in this regime. Nevertheless, both theory and observations suggest that GR can have significant classical and quantum corrections in the strong-gravity regime. 

With the direct detection of gravitational waves (GWs) in the ground-based detectors from such events, one should be able to test theories of gravity in the strong-gravity regime. As is always the case, any discovery serves as a springboard for subsequent discoveries by providing a new lens through which to view the Universe.
The signal-to-noise ratio (SNR) of the next generation of gravitational-wave detectors, such as Cosmic Explorer and Einstein telescope, will be at least 50 times greater than that of today's laser interferometer~\cite{2009-Sathyaprakash.Schutz-LRR,2013-Gair.etal-LRR,2015-Nakano.etal-PRD,2017-Abbott.etal-CQG}. These detectors are capable of confirming or refuting GR.

On the largest scales, the biggest surprise from observational cosmology is the accelerated expansion of the current Universe~\cite{1967-Sakharov-DANSF,2000-Sahni.Starobinsky-Int.J.Mod.Phys.D,2003-Padmanabhan-Phys.Rept.,2003-Peebles.Ratra-RMP}. This can be explained either by the presence of an exotic source referred to as dark energy or modifications to GR on the largest length scales~\cite{2016-Joyce.etal-Ann.Rev.Nucl.Part.Sci.}. Therefore, testing GR using cosmological and GW observations is crucial to understanding the Universe. With the new missions coming up, we are in a position to answer the above questions. 

First, however, to identify the classical and quantum corrections to GR in the strong and weak gravity regimes, we need to construct modified theories of gravity that make definite predictions. One direction to explore is to identify the core principles of GR, to which possible modifications can be introduced. Einstein used Mach's and Equivalence principles to formulate the general theory of relativity~\cite{1973-Misner.etal-Book,1984-Wald-Book,2014-Padmanabhan-GravitationFoundationsfrontiers}. Mach's principle suggested that local physical laws are affected by the large-scale structures of the Universe. Einstein later refined this as \emph{presence of matter curves the geometry of space-time}. Equivalence principle states that \emph{locally a free-falling observer and an inertial observer are indistinguishable}. As a consequence, inertial mass and gravitational mass are the same. In GR, the {four-dimensional (4D)} space-time and its properties are described by the metric tensor denoted by $g_{\mu \nu}$. The metric that determines the infinitesimal distance between two space-time points is described by the invariant quantity:
\begin{equation}
ds^2 = g_{\mu\nu} dx^\mu dx^\nu.
\end{equation}
Mathematical description of the relation between the geometry of the space-time and the distribution of the matter can be obtained from the Einstein-Hilbert action:
\begin{equation}
\mathcal{S}_{\rm EH} = \frac{1}{2 \kappa^2}\int d^4x \sqrt{-g} R + \int d^4x \sqrt{-g} {\cal L}_m
\label{eq:EHaction}
\end{equation}
where $R$ is the Ricci scalar, ${\cal L}_m$ is the matter Lagrangian and $\kappa^2 = 8 \pi G/c^4$. Varying the action leads to Einstein equations:
\begin{equation}
R_{\mu \nu} - \frac{1}{2} g_{\mu\nu} R  \equiv G_{\mu\nu} = \kappa^2 \, T_{\mu\nu}.
\label{eq:EinstEqu}
\end{equation}
The LHS describes the geometry of the space-time, while $T_{\mu \nu}$ represents the energy-momentum tensor of the minimally coupled matter field. 
Here $R_{\mu \nu} \equiv R^{\alpha}_{\mu \alpha \nu}$ denotes the Ricci tensor and the Riemann tensor is given by:
\begin{equation}
\label{def:Riemann}
R^{\alpha}_{\beta \gamma \delta}= \frac{\partial}{\partial x^{\g}} \Gamma^{\a}_{\b\d} 
- \frac{\partial}{\partial x^{\d}} \Gamma^{\a}_{\b\g} 
+ \Gamma^{\a}_{\g\epsilon} \Gamma^{\epsilon}_{\b\d} 
- \Gamma^{\a}_{\d\epsilon} \Gamma^{\epsilon}_{\b\g}.
\end{equation}
where
\begin{equation}
\label{def:Gamma}
\Gamma^\mu_{\alpha\beta} =\frac{g^{\mu\gamma}}{2}(g_{\gamma\alpha,\beta}+g_{\gamma \beta,\alpha}-g_{\alpha\beta,\gamma})
\end{equation}
and $g_{\gamma\alpha,\beta}$ refers to the partial derivative of $g_{\gamma\alpha}$
wrt $x^{\beta}$. 
In GR, the presence of the space-time curvature prevents two parallel geodesics from remaining parallel. Therefore, the relative acceleration between any two geodesics is described by the geodesic deviation equation:
\begin{equation}
\label{def:GeodDevi}
\frac{d^2 \xi^{\alpha}}{dt^2} = R^{\alpha}_{\mu\beta\nu} \frac{dx^{\mu}}{dt}
\frac{dx^{\beta}}{dt} \xi^{\nu}
\end{equation}
\begin{sidewaystable}
\hspace{-1.5cm}
\begin{tabular}{|c|c|} \hline
{ Principles/Predictions}&{\bf Observational tests} \\ \hline
\multirow{2}{*}{ Weak equivalence principle }& All freely falling test particles fall at the same rate irrespective of their internal composition. \\
 & [Schlamminger- 2008]~\cite{2008-Schlamminger.etal-Phys.Rev.Lett.} \\
 \hline
\multirow{2}{*}{ Local Lorentz invariance}& Non-gravitational physical laws are independent of the velocity of the freely falling frame described.  \\
 & [Rossi \& Hall-1941]~\cite{1941-Rossi.Hall-Phys.Rev.} \\
  \hline
\multirow{2}{*}{ Local position invariance}& Non-gravitaional physical laws are independent of the position in space or time of the \\
 & freely falling frame described. [Fischer et al.-2004]~\cite{2004-Fischer.others-Phys.Rev.Lett.}\\ \hline
\multirow{2}{*}{ Gravitational deflection of light}& A ray of light passing by a massive object will bend towards the object, changing the apparent \\
 &  position of the source from which it was emitted  [van der Wel et al.-2013]~\cite{2013-Wel.others-Astrophys.J.Lett.}\\ \hline
 \multirow{2}{*}{ Perihelion precession of Mercury}&Unexplainable by Newtonian gravity, the slow perihelion shift of Mercury at 43 arcseconds per \\& century was correctly predicted by GR.  Most recent observation include Messenger  \\& spacecraft observations [Park et al.-2017]~\cite{2017-Park.etal-TheAstronomicalJournal}\\ \hline
  \multirow{2}{*}{ Lense-Thirring effect}& A rotating massive body will drag inertial frames around its vicinity along the direction of rotation,\\& leading to a precession in a gyroscope's spin if it is not parallel to the angular momentum \\& of the rotating body. [Ciufolini et al.-2016]~\cite{2016-Ciufolini.others-Eur.Phys.J.C}\\ \hline
  \multirow{2}{*}{ Gravitational redshift}&An observer located at a distance from a massive luminous body will see the light from the body to \\& be redshifted compared to an observer located closer to the surface of the body.  
	[Barstow et al.-2005]~\cite{2005-Barstow.etal-Mon.Not.Roy.Astron.Soc.}\\ \hline
  \multirow{2}{*}{ Gravitational waves}& Mass distributions with time-varying Quadrupole moments and higher will shed orbital/rotational  \\& energy by radiating GWs. [Abbott et al.-2015]~\cite{2016-Abbott.Others-Phys.Rev.Lett.}\\ \hline
\multirow{2}{*}{ Existence of BHs}& Direct imaging through the Event Horizon telescope of the central BH of neighboring galaxy M87  \\
 & [Akiyama et al.-2019]~\cite{2019-Akiyama.others-Astrophys.J.Lett.} \\ \hline

\end{tabular}
\caption{List of the tests of the principles and predictions of GR.}
\label{tab::grsuccess}
\end{sidewaystable}
As discussed above, Einstein employed aesthetic and philosophical criteria in developing GR, allowing the development of alternative theories based on diverse criteria. Such efforts to obtain alternative theories to GR began as early as the 1920s, with examples including Whitehead's quasi-linear theory, Einstein's Carton theory, and Fierz-Pauli theory~(See, for instance, \cite{Willbook}). However, there was no reason to question GR for many decades as it was self-contained and successful in explaining experimental and observational results. Table~\ref{tab::grsuccess} contains the predictions and principles of GR that observations and experiments have verified~\cite{2019-Bhattacharyya-Phd}.

However, as we discuss in the next section, both theory and observations suggest that
GR might have significant corrections in the strong and weak gravity regimes. This review covers three aspects of modified theories of gravity: Why, How, and What.
The title of this review is adopted from Simon Sinek's Golden circle~\cite{2009-Sinek-Why}. Sinek proposed the concept of Golden Circle to explain how certain brands/companies can inspire and truly differentiate from others. In the same spirit, {MG} theories must have a theoretical/ observational reason of:
\begin{figure}[!h]
\includegraphics[width=0.5\textwidth]{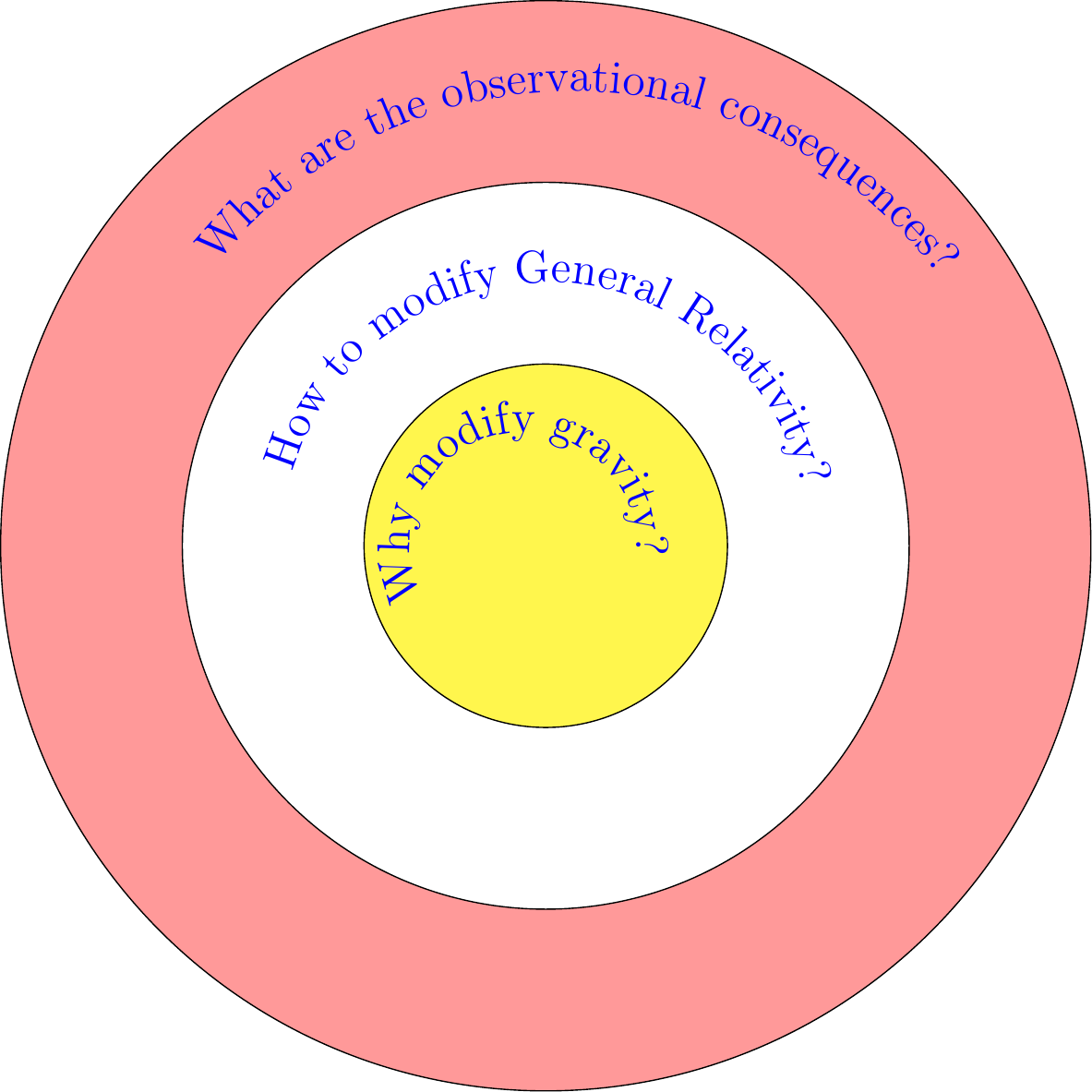}
\end{figure}
(i) Why do we need to consider them? (ii) How can they be physical? and (iii) What are the distinguishing features? This review covers these three aspects of {MG} models. For lack of space, this review is not exhaustive, although it accurately reflects the current literature's state and illustrates the complexity of building an alternative to GR. For other aspects of {MG} theories, readers can see Refs.~\cite{2009-Alexander.Yunes-PRep,2010-DeFelice.Tsujikawa-LivingRev.Rel.,2010-Sotiriou.Faraoni-Rev.Mod.Phys.,2011-Capozziello.DeLaurentis-Phys.Rept.,2011-Hinterbichler-RMP,2012-Clifton.etal-Phys.Rept.,2014-deRham-LRR,2015-Joyce.etal-Phys.Rept.,2017-Nojiri.etal-Phys.Rept.}. We use $(-,+,+,+)$ metric signature, and $8 \pi G = \kappa^2 = 1/\MPl^2 \sim \ell_P^2$ where the Planck Mass $(\MPl) \sim 10^{19}~{\rm GeV}$. Unless otherwise specified, we set $c = \hbar =1$.

\section{Why modify GR?}
\label{sec3}
Einstein equations \eqref{eq:EinstEqu} are obtained by varying the Einstein-Hilbert action \eqref{eq:EHaction} which is linearly dependent on the curvature scalar $R$. 
In this section, we list four reasons why the assumption of the linear dependence of curvature scalar might not be valid.  

\subsection{Gravity is not probed at all scales} 

As mentioned in the Introduction, one of GR's founding principles is Einstein's equivalence principle (EEP), which states that all non-gravitational phenomena are locally unaffected by gravity when performed in a freely falling frame. A particular consequence of this principle is that everything, including light, obeys the same laws. Thus, the EEP represents the interface between gravity and the rest of physics. However, unlike gauge invariance of electromagnetic field, it is not a fundamental symmetry but an experimental fact. Einstein initially referred to it as the equivalence hypothesis before elevating it to a principle once it became clear how critical it was to generalize special relativity to include gravitation. Indeed, it is surprising that the EEP is satisfied, let alone under the stringent uncertainties of modern tests.

Gravity is well tested in solar and stellar system scales and is ill-tested beyond this scale~\cite{Willbook,2006-Will-LivingRev.Rel.}. One way to test GR is to test the underlying principle of GR --- the equivalence principle. Hence, any experimental evidence of the violation of the equivalence principle will also act as evidence against GR. Additionally, one of its components, local Lorentz invariance, implies {charge, parity, and time reversal (CPT) symmetry}~\cite{1951-Schwinger-PR}, which is well tested on Earth~\cite{2013-Liberati-CQG}. Nonetheless, testing the EEP's validity on cosmic scales is much more difficult.

\begin{figure}
\includegraphics[width=0.5\textwidth]{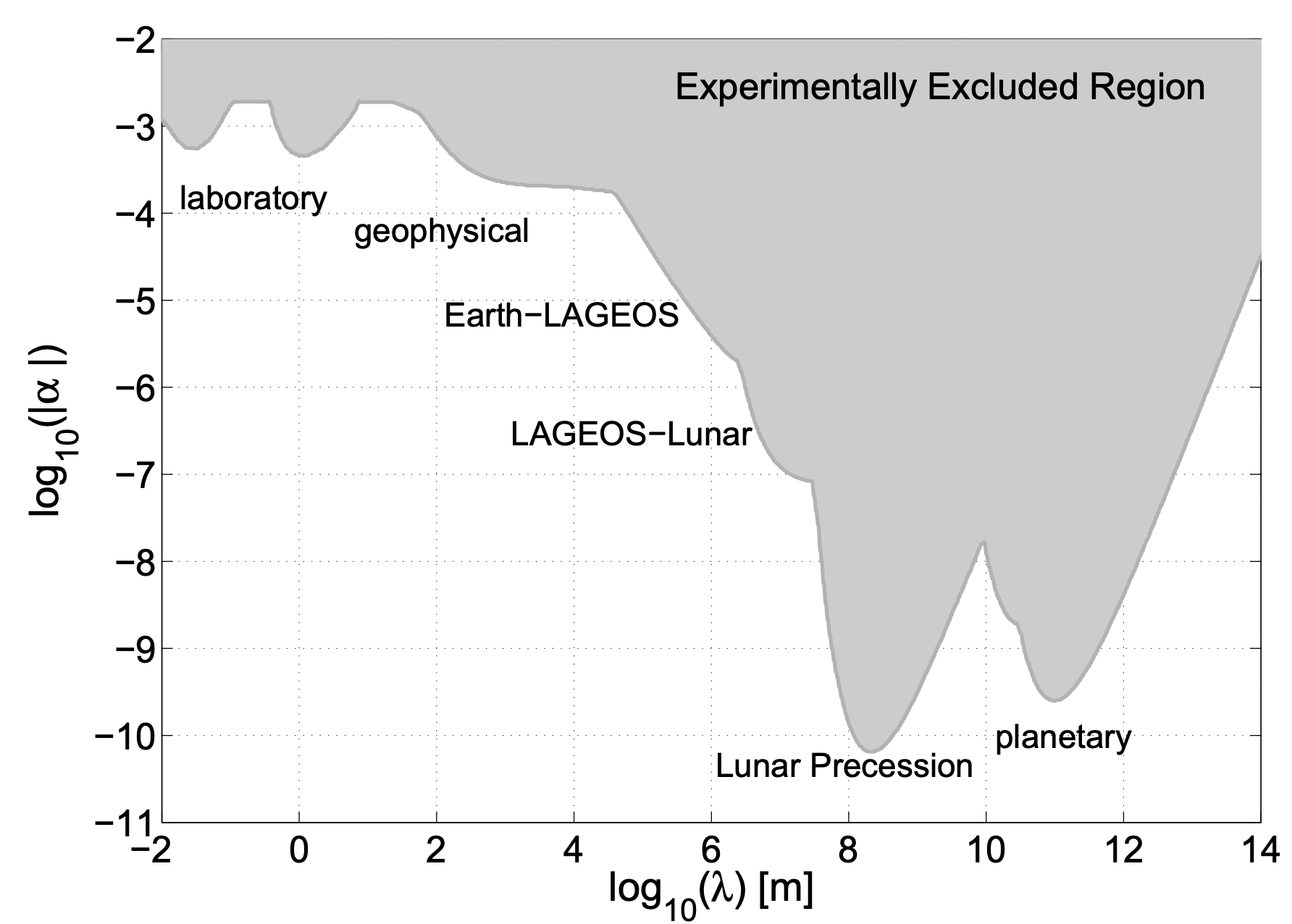}
\caption{Plot shows the current 95\% confidence-level constraints on the inverse-square-law-violating Yukawa interactions 
\eqref{eq:yukawa} with $\lambda  > 1 {\rm cm}$~\cite{2017-Berge}.} 
\label{fig:Yukawa}
\end{figure}

\begin{figure}[!h]
\includegraphics[width=0.7\textwidth]{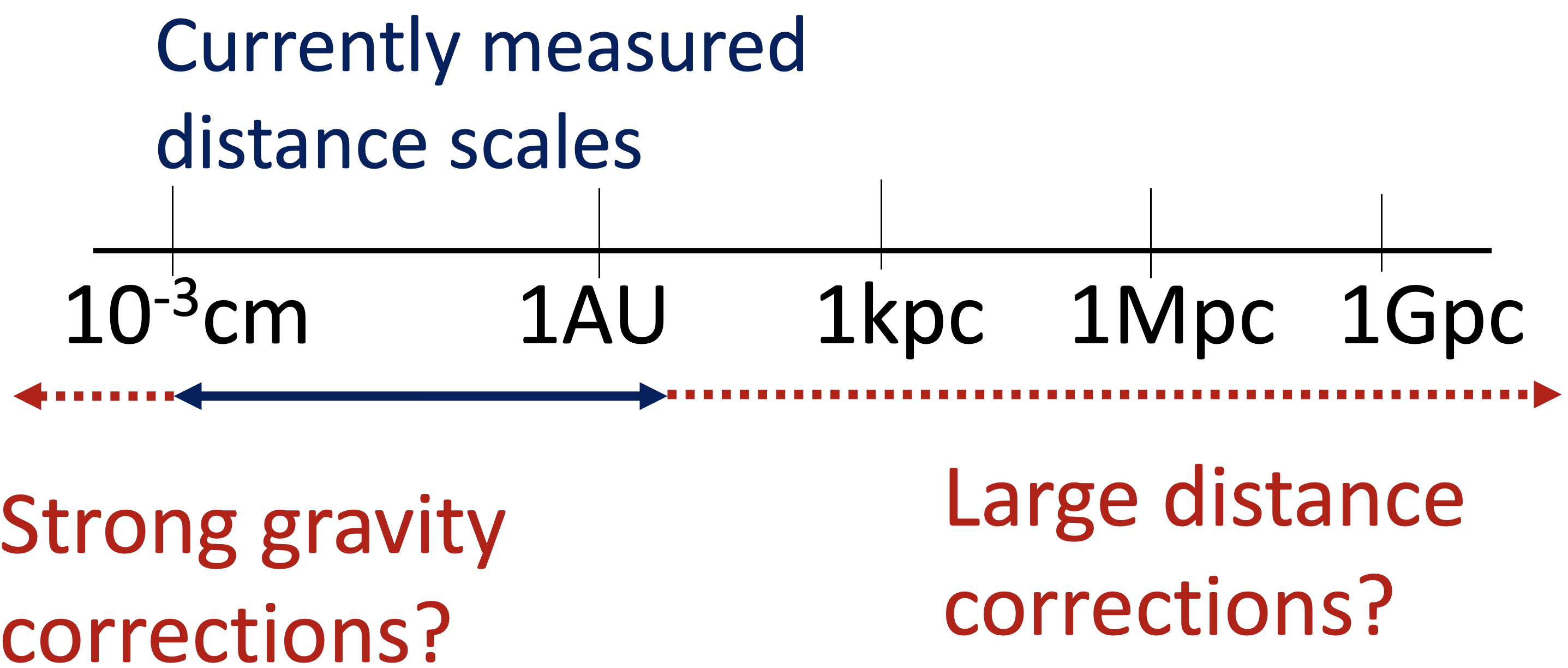}
\caption{This figure provides a representation of the range of length/scale where GR has currently been tested.}
\label{fig:MGtested}
\end{figure}

In principle, other gravitational fields besides the metric could exist, such as scalar fields. Thus, the existence of a light or massive scalar field with a coupling to matter weaker than gravitational
strength is a possible source of violation of the weak equivalence principle. In the weak-field limit, GR leads to Newtonian gravity with inverse square law. Fig. \ref{fig:Yukawa} shows the experimentally excluded region for inverse-square-law-violating Yukawa type gravity interaction~\cite{2017-Berge,2015-Murata.Tanaka-CQG}:
\begin{equation}
V(r) = -G \frac{m M}{r}\left[1+\alpha e^{-r / \lambda}\right]
\label{eq:yukawa}
\end{equation}
where $\alpha$ is a dimensionless constant and $\lambda$ corresponds to the deviation of the Newtonian potential.
Thus from the above plot, we can conclude that the inverse-square-law predicted by GR may not be conclusive for all scales. Fig. \eqref{fig:MGtested} provides a representation of the range in which gravity has been tested accurately. With the availability of improved observational/experimental data, we might have to consider strong gravity corrections or large distance corrections depending on the length/energy scales. In the rest of the section, we provide reasons for the modifications to GR in strong gravity and at the largest possible distances. 

\subsection{GR needs corrections in strong gravity regimes} 

In the earlier subsection, we showed that the gravity is not well tested beyond certain scales. We have extrapolated GR from the current observational scales to strong gravity regimes. 
Considering two specific cases, we show that 
the very success requires modifications to GR at strong gravity regimes.

First, let us consider the Friedmann-Lemaitre-Robertson-Walker (FLRW) line element:
\begin{equation}
ds^2 = - dt^2 + {a^2(t)  d\bar{x}^2}  
 = -a^2(\eta) \left[d\eta^2 -  d\bar{x}^2\right]
 \label{eq:FRW}
\end{equation}
where $t$ is the cosmic time, $\bar{x}$ refers to the 3-space, and $\eta$ is the conformal time. Assuming a power-law scale factor $a(t) \sim t^n$ ($n>0$) that is a solution to Einsten's equation \eqref{eq:EinstEqu}, the Ricci scalar for this line-element is:
\begin{equation}
 R \sim \dfrac{n(2n-1)\MPl^2}{\tau^2}, \qquad \tau=t \, M_{Pl}
\end{equation}
Let us consider the following series:
\begin{equation}
f(R)=R+\sum_{m \geq 1}\dfrac{\alpha_m}{\MPl^{2m}} R^{m+1}
\end{equation}
where $\alpha_m$ are dimensionless constants. Ratio between $(m+1)^{th}$ term and $m^{th}$ term in the above series is:
\begin{equation*}
 \dfrac{f_{m+1}}{f_{m}} \sim \dfrac{n(2n-1)}{\tau^2} \qquad \mbox{for all}~m
  \end{equation*}
From the above expression, we infer that higher-order Ricci scalar $R$ will significantly contribute at early times (small $\tau$) or small length scales. Thus, the very success requires modifications to GR, and it is imperative to include 
higher-order curvature terms at early times.

To further emphasize this, we consider Schwarzschild space-time which is an exact solution of Einstein's equation:
\begin{equation}
ds^2 = - \left(1 - \frac{r_H}{r} \right) dt^2 +  \left(1 - \frac{r_H}{r} \right)^{-1} \!\!\! dr^2 + r^2 \left( d\theta^2 + \sin^2\theta d\phi^2 \right), 
\quad \mbox{where}~r_H~\mbox{is horizon radius}
\label{eq:SchwST}
\end{equation}
Although Schwarzschild space-time has a vanishing Ricci tensor (scalar), not all the components of the Riemann tensor vanish. From the Riemann tensor, we can construct Kretschmann scalar:
\begin{equation}
K(r) = R_{\mu\nu\alpha\beta} R^{\mu\nu\alpha\beta} = 
\frac{12 r_H^2}{r^6},~~\mbox{where}~~r_H = \frac{2 G M}{c^2}
\end{equation}
which is non-zero. The table below gives the value of the Kretschmann scalar at the horizon radius $r_H$ for different BH masses:

\begin{table}[!h]
\begin{tabular}{|c|c|c|}
\hline
& &  \\[-1.3em]
Black hole &  $r_H$  (m) & $\sqrt{K(r_H)}$ (m$^{-2}$)\\
\hline
& &  \\[-0.8em]
Solar Mass & $10^{3}$  & $10^{-6}$   \\
& &  \\[-1.8em]
\hline 
& &  \\[-0.8em]
PBH of $10^{-3} M_{\odot}$  & $1$ & $3$  \\
& &  \\[-1.8em]
\hline 
& &  \\[-0.8em]
PBH of $10^{-5} M_{\odot}$  &  $10^{-2}$ &  $10^4$\\
& &  \\[-1.8em]
\hline 
& &  \\[-0.8em]
PBH of $10^{-10} M_{\odot}$ & $10^{-7}$ & $10^{14}$  \\
\hline
\end{tabular}
\caption{{Value of the Kretschmann scalar at the horizon radius $r_H$ for different BH masses}}
\label{tab::BH_Kr_scalar}
\end{table}

From the above table, we infer that the description of  Primordial black holes (PBHs)~\cite{1971-Hawking-MNRAS,1979-Zeldovich_Novikov-NASA,1975-Chapline-Nature} may require one to go beyond GR. 
{In order to demonstrate this, let us consider the following MG Lagrangian: }
\begin{equation}
{\mathcal{L}_{PBH} = R + \alpha R^{2}  + \beta R^{\mu \nu}R_{\mu \nu}  + \gamma R^{\mu \nu \sigma \delta}R_{\mu \nu \sigma \delta}}
\end{equation}
{where $\alpha$, $\beta$ and $\gamma$ are arbitrary constants with the dimensions $\left[ L^2 \right]$. Consider a BH described by Schwarzchild space-time~\eqref{eq:SchwST}. Although the Ricci scalar ($R$) and Ricci tensor ($R_{\mu \nu}$) vanish, Riemann curvature is non-zero. Thus in the above Lagrangian, the Kreschtmann scalar is non-zero. As we see in Table~\ref{tab::BH_Kr_scalar}, even though this contribution is negligible in the case of solar mass BH, for PBH with lower mass, we cannot ignore the effect of the Kretschmann scalar. More importantly, we need to consider higher curvature terms to understand the physics of PBHs. The same reasoning also applies to Kerr BHs, which has vanishing Ricci tensor.}

From these two physical situations, one can conclude that the form of classical GR ~\eqref{eq:EHaction} does not reflect nature, and indeed one would at the very least expect higher-order curvature terms that induce alterations to GR. Note that these corrections can be classical or quantum. 

\subsection{Late time acceleration and new physics}

The Universe is homogeneous and isotropic in the cosmological scales and is described by the FLRW metric \eqref{eq:FRW}.
The $\Lambda$CDM model is considered to be the standard model of cosmology and is consistent with most of the observations ~\cite{2000-Padmanabhan-TheoreticalAstrophysicsVolume}. 
The $\Lambda$CDM model presupposes that GR describes gravity in the cosmological scales. 
As the name implies, the model postulates that dark matter and dark energy denoted by the cosmological constant $\Lambda$ dominate the Universe's energy budget.

The success of the standard model of cosmology comes at a price as atomic matter makes up less than 5\% of our Universe. Additionally, we do not know what dark energy and dark matter are made of. Dark energy and dark matter have not been directly detected yet other than inferring through the gravitational interaction. $\Lambda$CDM model is also faced with several theoretical and observational challenges.
A few of those problems are given below.

\noindent \textit{Fine tuning problem}: 
From the observations, it is estimated that the current energy density of the cosmological constant is of the order of $\rho_{\Lambda}\sim 10^{-47}GeV$~\cite{2000-Sahni.Starobinsky-Int.J.Mod.Phys.D,2003-Padmanabhan-Phys.Rept.,2006-Copeland.etal-Int.J.Mod.Phys.}.
 Assuming the cosmological constant originates from a vacuum energy density, this value is in severe disagreement with the value of vacuum energy density predicted by the quantum field theory $\rho_{vac} \sim 10^{74} GeV$~\cite{1989-Weinberg-Rev.Mod.Phys.}. In other words, the value of $\rho_{\Lambda}$ is in conflict with the possible energy scales and requires fine-tuning.

\noindent \textit{Coincidence problem}: Cosmological observations suggests that the current value of dark energy density parameter to be $\Omega^{(0)}_{\Lambda} \sim 0.7$, which is of the same order of the current value of the matter-energy density parameter given by $\Omega^{(0)}_{m} \sim 0.3$. In other words, even though the matter-energy density $\rho_m$ changes with time and the dark energy density $\rho_{\Lambda}$ remains constant, they are of the same order precisely at the current epoch, which appears to be a coincidence, and would require fine-tuning of parameters in the early Universe. This is known as the cosmological coincidence problem~\cite{2006-Copeland.etal-Int.J.Mod.Phys.}. It has to be noted that this problem appears in many other dark energy models as well~\cite{2015-Amendola.Tsujikawa-DarkEnergyTheory}.

\noindent \textit{Tensions between early and late universe observations}: As mentioned earlier, the $\Lambda$CDM model is consistent with the cosmological observations. This includes the observations from the Cosmic Microwave Background (CMB) and the late Universe (Late time acceleration). However, the latest observations suggest that within the framework of $\Lambda$CDM, there are some tensions between these two sets of observations~\cite{2019-Verde.etal-NatureAstron.}.

The estimated value of $H_0$ from the observations of the late Universe shows $5\sigma$ tension with the value of $H_0$ estimated from the CMB observations~\cite{2021-Riess.others}. Early Universe estimates of $H_0$ result in higher values as compared to the ones obtained from the local Universe observations. With the increasing accuracy of the observations, it has been suggested that one might have to look beyond the $\Lambda$CDM model to alleviate the $H_0$ tension.

Another prominent cosmological tension comes from the observations of large-scale structures and the corresponding parameter values estimated from CMB observations. This tension ($3\sigma$) can be seen in the parameter constraints on the matter-energy density parameter $\Omega_m$, the amplitude $\sigma_8$ of matter fluctuations estimated from the local Universe measurements and CMB observations~\cite{2013-Macaulay.etal-Phys.Rev.Lett.,2021-Asgari.others-Astron.Astrophys.}.

Various alternate theories of gravity have been proposed to solve these issues ~\cite{2012-Clifton.etal-Phys.Rept.,2021-DiValentino.etal-Class.Quant.Grav.}. These range from {MG} models with geometric modifications~\cite{2017-Nojiri.etal-Phys.Rept.} to scalar-tensor theories ~\cite{2006-Copeland.etal-Int.J.Mod.Phys.}. It has been shown that these models can drive the observed late-time evolution of the Universe. However, most of these cosmological models are not preferred over the $\Lambda$CDM model by the cosmological observations. Hence, it is important to construct alternatives to the $\Lambda$CDM model that better fits the observational data.

\subsection{Constructing quantum theory of gravity}

According to our current understanding, the fundamental interactions of nature are the strong, electromagnetic, weak, and gravitational interactions. The first three are successfully explained by standard model of particle physics, which partially unifies the electromagnetic and weak interactions. Except for the non-vanishing neutrino masses, no observable data contradicts the standard model at the moment. GR describes gravity. Thus, there is no incentive to seek out new physical rules from an empirical standpoint. However, the situation is unsatisfactory from a theoretical (mathematical and conceptual) standpoint. GR is a classical theory, whereas the standard model is a quantum field theory that describes an incomplete unification of interactions. 

We still do not have a physically and mathematically consistent theory of quantum gravity. As we go to smaller and smaller length scales, the quantum effects cannot be ignored. For example, consider a particle of mass $M$. Quantum mechanics gives a strict lower  bound on the length (Compton wavelength) within which $M$ can be localized:
\begin{equation}
\lambda_C = \frac{h}{M c} \quad \Longrightarrow \quad \lambda_C \downarrow~\mbox{as}~M \uparrow.
\end{equation}  
According to GR, any particle of mass $M$ can become a BH if it collapses down to the Schwarzschild radius: 
\begin{equation}
 r_H = \frac{2 G M}{c^2} \quad \Longrightarrow \quad  
r_H  \uparrow~\mbox{as}~M \uparrow
 \end{equation}
Note that $\lambda_C$ decreases as the Mass \red{$M$} increases while $r_H$ increases as the Mass increases. It thus follows that for Planck mass $M_{\rm Pl}$ size BHs, Planck length $l_{\rm Pl}$, $r_H$ and $\lambda_C$ are comparable. At such small length scales, 
quantum gravitational effects become important.
Thus, quantum gravitational effects are expected (when considered at low energy) to correct the classical action \eqref{eq:EHaction} by additional terms involving quadratic and higher powers of the curvature tensor~(for a historical overview, see Ref. \cite{2007-Schmidt-IJGMMP}). 

Considering quantum matter in a classical gravitational background~\cite{1982-Birrell.Davies-Quantumfieldsin} presents additional intriguing issues, most notably the possibility that the zero-point fluctuations of the matter fields generate a non-vanishing vacuum energy density $\rho_{\rm vac}$. This is analogous to adding the term $\Lambda g_{\mu\nu}$ to the left-hand side of Einstein's equations \eqref{eq:EinstEqu}, where $\Lambda = 8G \rho_{\rm vac}/c^4$. As mentioned in the previous subsection, recent cosmological observations imply a positive cosmological constant corresponding to $\rho_{\rm vac} \sim (2.3 \times 10^{-3} {\rm eV})^4$.
In this {review}, we only consider the classical limit of gravitation (i.e., classical matter and classical gravity). 


\section{How to modify GR?}\label{sec4}

In the preceding section, we discussed various examples that demonstrate the need to modify GR to describe gravity at all length scales accurately. Ironically, the very success of GR necessitates its modification. However, due to the complexity and diversity of the field of the modification of GR, we must first understand what makes GR unique. Understanding this will assist in determining whether it is feasible to construct alternate theories of gravity and, if so, how to proceed.

To accomplish this, we will examine Lovelock's theorem~\cite{1971-Lovelock-JMP,1972-Lovelock-JMP}. While the Lovelock theorem is a no-go theorem, it does provide direction for searching for modifications to GR in four dimensions and beyond. We begin by stating the Lovelock theorem and the five conditions it imposes on GR being the unique theory of gravity in four dimensions. We then demonstrate that when each of these conditions is relaxed, paths to construct consistent modified theories of gravity become available.

\subsection{Is GR unique in {4D}?}

Lovelock's theorem~\cite{1972-Lovelock-JMP} states that Einstein {field} equations~\eqref{eq:EinstEqu} are \emph{the only} second-order local equations of motion for a metric derivable from the action in {4D}. In other words, Lovelock's theorem proves that GR is unique under the following five conditions:
\begin{enumerate}
\item Equations of motion {(EOM)} should be second order (quasi-linear). 
\item No extra fields (degrees of freedom {(DOF)}).
\item Covariance $\partial_{\mu} \to \nabla_{\mu}$.
\item Locality.
\item Space-time is 4D.
\end{enumerate}
Lovelock's theorem acts as a no-go theorem for the gravity theories for the {4D} space-time. This implies that any gravitational theory other than GR is ruled out under the aforementioned conditions. However, the preceding five conditions suggest possible ways around Lovelock's theorem. To begin, we determine whether GR is unique in higher dimensions.

\subsection{Is GR unique in higher dimensions?}
Lovelock obtained the most general second rank tensor in arbitrary D-dimensions that satisfy the following three conditions~\cite{1971-Lovelock-JMP}:
(1) Symmetric 
(2) Depends on metric and \red{its derivatives up to II order} {its first and second order derivatives} 
(3) Divergence-free.
Lovelock showed that the Einstein-Hilbert action \eqref{eq:EHaction} is no longer unique. 
The unique, D-dimensional theory with the same properties contains 
nonlinear corrections to the Einstein-Hilbert Lagrangian, and these terms reduce to Einstein theory in {4D} space-times~\cite{2010-Dadhich-Pramana,2013-Padmanabhan.Kothawala-PRep}. Table~\ref{tab::highD_lag} gives the unique gravity Lagrangian for higher-dimensional space-times. As can be seen, for $D > 4$, \red{Einstein gravity} {GR} can be thought of as a particular case of Lovelock gravity since the Einstein-Hilbert term is one of several terms that constitute the Lovelock action. The Lovelock theories are free from ghosts. Interestingly, gravity may travel faster or slower than light~\cite{2014-Reall.etal-CQG}.

\begin{table}[!hbt]
\begin{tabular}{|c|c|}
\hline
&  \\[-1.3em]
Space-time  &  Unique gravity Lagrangian \\
\hline
&  \\[-0.8em]
D = 3, 4  & $L_{\rm {I}} = R + \Lambda$  \\
&   \\[-0.8em]
\hline 
&   \\[-0.8em]
D = 5, 6  & $L_{\rm {II}} = L_{\rm {I}} + \alpha_{\rm GB} L_{\rm GB}$ \qquad
$L_{\rm GB} = R_{\alpha \beta \gamma \delta} R^{\alpha \beta \gamma \delta}-4 R_{\alpha \beta} R^{\alpha \beta}+R^{2} $ \\
&   \\[-0.8em]
\hline 
&   \\[-0.8em]
D = 7, 8   &  ${\small \begin{aligned}
L_{\rm {III}}&= L_{\rm {II}} + R^{3}+3 R R^{\mu \nu \alpha \beta} R_{\alpha \beta \mu \nu}-12 R R^{\mu \nu} R_{\mu \nu}+24 R^{\mu \nu \alpha \beta} R_{\alpha \mu} R_{\beta \nu}  \\
&+16 R^{\mu \nu} R_{\nu \alpha} R_{\mu}^{\alpha} +24 R^{\mu \nu \alpha \beta} R_{\alpha \beta \nu \rho} R_{\mu}^{\rho}+8 R^{\mu \nu}{ }_{\alpha \rho} R_{\nu \sigma}^{\alpha \beta} R_{\mu \beta}^{\rho \sigma}+2 R_{\alpha \beta \rho \sigma} R^{\mu \nu \alpha \beta} R^{\rho \sigma}{ }_{\mu \nu}
\end{aligned}}$ \\
\hline
\end{tabular}
\caption{Unique gravity Lagrangians in higher dimensional space-times.}
\label{tab::highD_lag}
\end{table}

\subsection{Bypassing Lovelock theorem in {4D} and possible modifications}

We showed how the Lovelock theorem~\cite{1972-Lovelock-JMP} provided a route to obtain higher-dimensional gravity action consisting of the dimensionally-extended Euler densities 
(polynomial scalar densities in the Riemann curvature tensor with the property that their Euler-Lagrange derivatives contain derivatives of the metric only up to second order). Similarly, as listed in the table below, each of the remaining four conditions provides options to bypass Lovelock's theory in {4D}, leading to a specific modification to GR. 
\begin{table}[!htb]
\begin{tabular}{|c|c|}
\hline
{\bf Lovelock conditions}  &  
{\bf By-pass for {MG} theories} \\ 
\hline
&  \\[-1.2em]
Second order EOM & Beyond second order EOM  \\
&   \\[-1.8em]
\hline 
&   \\[-1.2em]
No extra fields & Add new field content (Scalar, Vector, Tensor) \\
&   \\[-1.8em]
\hline 
&   \\[-1.2em]
Covariance   &  Non-minimal coupling to matter fields \\
&   \\[-1.8em]
\hline 
&   \\[-1.2em]
Locality & Non-local theories \\ 
\hline 
&   \\[-1.8em]
{4D} & Higher dimensions (5D and above) \\
\hline
\end{tabular}
\caption{{By-passing Lovelock theorem for modified gravity theories}}
\label{tab::bypass_lovelock}
\end{table}
\begin{figure}
\center{
\includegraphics[scale=1.25]{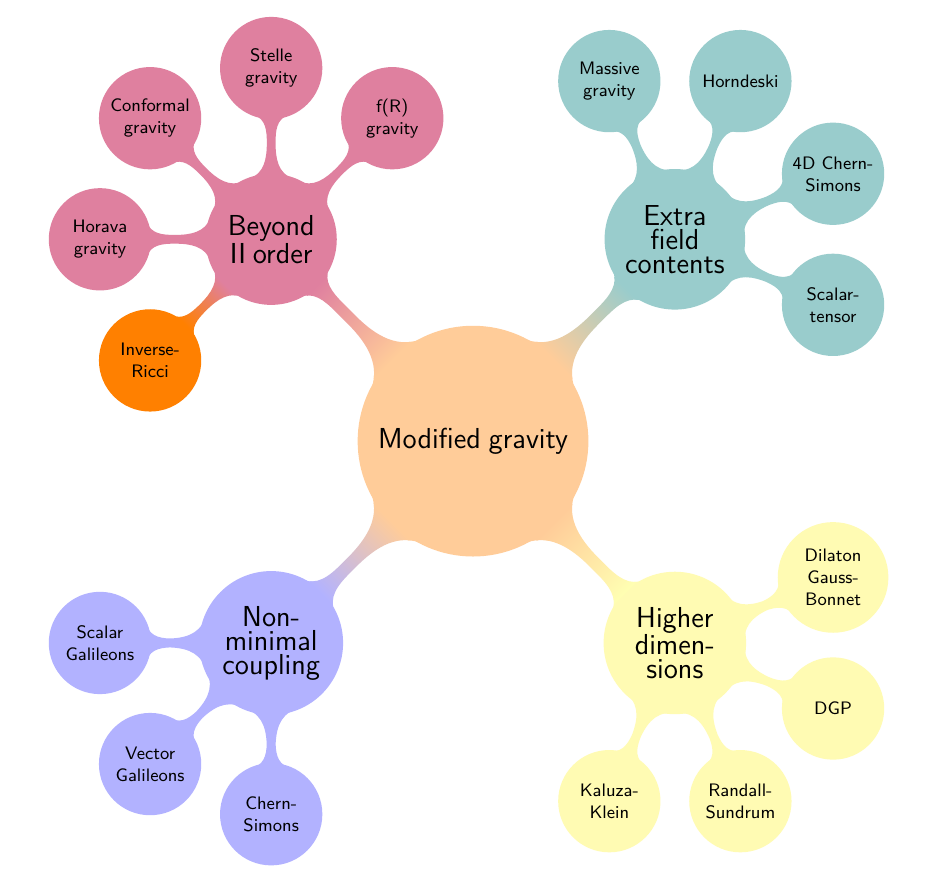}
}
\caption{Classification of modified gravity theories}
\label{fig:mgoption}
\end{figure}
From Table~\ref{tab::bypass_lovelock}, we can classify the {MG} models into four broad classes: 
\begin{enumerate}
    \item Beyond second order 
    \item Extra field contents
    \item Non-minimal coupling
    \item Higher dimensions
\end{enumerate}
Several models have been proposed under each of the class of MG theories. The mind map in Fig.~\ref{fig:mgoption}  provides a bird's eye view of various MG models proposed in the literature. A few points are in order: First, some models, like Inverse-Ricci~\cite{2020-Amendola.etal-Phys.Lett.B,2022-Das.etal} can have a combination of two or more of these types of modifications. Second, we have included a few generic models that are theoretically possible. Some models have been ruled out due to theoretical and observational inconsistencies. Third, not all modifications are equally important for MG model building. Hence, we have not included non-local {MG} theories in this review.

\section{Constructing modified theories of gravity}

From the above discussions, we can say that, like the
standard model of particle physics,
GR is a fundamental theory and is an effective description valid in a certain energy range. The first step in constructing a consistent gravity theory requires us to list the criteria for such theories~\cite{2006-Will-LivingRev.Rel.}. 
We then discuss the criteria in the effective field theory language~\cite{2007-Burgess-ARNPS,2012-Donoghue.etal-Talk,2015-Donoghue.Holstein-JPG,2019-Ruhdorfer.etal-JHEP,2020-Penco-Arxiv}.

\subsection{Criteria for a consistent gravity theory}
Will has provided the following list of criteria that any consistent gravity theory must satisfy~\cite{Willbook}:
\begin{enumerate}
\item \textbf{Theory must be complete:}~
The theory should be able to analyze from {\sl first principles} the outcome of any experiment.

\item \textbf{Theory must be self-consistent:}~
Predictions should be unique and independent of the calculation method. {In other words, the theory should be able to make predictions that can be tested by observations/experiments, making it falsifiable.} 

\item \textbf{Theory must be relativistic:}~ 
Special theory of relativity has been shown to be consistent with several experimental results. It governs the kinematics and dynamics in a flat space-time. Hence any theory of gravity should reduce to Special Relativity when gravity is \emph{turned off}. This requirement is also important for the development of quantum theory of gravity, since existing theories of quantum physics are constructed to be consistent with the special theory of relativity.

\item \textbf{Theory must have the correct Newtonian limit:}~Before GR, Newton's theory of gravity was considered to be the standard description of gravity. It has been consistent with various terrestrial and solar system experiments and observations indicating that in most of the weak field and non-relativistic situations, Newton's theory of gravity is valid. Hence any theory of gravity should reproduce Newton's laws in the weak gravitational fields and slow motion.
\end{enumerate}
We now translate these criteria in the effective field 
theory language.

\subsection{Translating the criteria in the language of effective field theory}

Effective field theories are, by definition, quantum field theories based on the principles of relativity and quantum mechanics but without the restriction of renormalizability. Although our analysis is strictly classical, the effective field theory language is useful for comprehending the potential issues we may face while constructing {MG} theories. Like in any effective theory, any {MG} theory must have the following four ingredients: 
\begin{itemize}
\item \textbf{Decoupling:} Details of very small-distance phenomena are largely irrelevant for the description of much larger systems. In this regard, the strong gravity corrections to GR and modifications to GR in the cosmological scales should not be decoupled. For instance, the Horava-Lifshitz model has strong coupling problems, extending all the way into the deep infra-red~\cite{2009-Charmousis.etal-JHEP}. 
$1/R$ gravity, which can explain the current acceleration of the Universe~\cite{2004-Carroll.etal-PRD}, suffers from violent instabilities and is inconsistent with solar-system test~\cite{2003-Dolgov.Kawasaki-PLB,2006-Erickcek.etal-PRD}. Thus, any {MG} model with the correct Newtonian limit should have clear separate energy scales. 

\item  \textbf{Degrees of Freedom:} We need to know what DOF is relevant to describe the physical system we are interested in. Once the relevant degrees of freedom are identified, the description of the natural phenomena can be much simpler. In the case of GR, the degrees of freedom are determined by the independent components of the metric tensor; however, for {MG} theories, there can be extra degrees of freedom like the scalar degree of freedom in scalar-tensor theories. However, for the modified theory to be complete and self-consistent, the number of degrees of freedom must be unique and make testable predictions. Hence, in any scalar-tensor theories of gravity, the number of scalar fields is usually one or two, and these decouple at low-energies~\cite{2010-DeFelice.Tsujikawa-LivingRev.Rel.}  

\item \textbf{Expansion parameters:}  In principle, action can have infinite terms. Therefore, we expand the action with one or more expansion parameters. For example, the Einstein-Hilbert action~\eqref{eq:EHaction} has a linear dependence on the curvature scalar. As discussed earlier, we need to include higher-order terms while considering strong gravity. 

\item \textbf{Symmetries:} Identify the symmetries that describe the system at a given energy or length scale. GR is diffeomorphism invariant. Thus, any modified theory of gravity must preserve this and be locally Lorentz-invariant at the low-energy limit.
\end{itemize}

At energies lower than the Planck scales, Donoghue has shown the quantum theory of GR  exists and is of the form of an effective field theory~\cite{2012-Donoghue.etal-Talk,2015-Donoghue.Holstein-JPG}. More specifically, he has shown systematically that: (i) Higher-order loops are made finite by counterterms provided by the higher-order terms in the Lagrangian. (ii) The non-renormalizability of GR is not a problem; GR can be renormalized perturbatively. (iii) Quantum GR is an excellent perturbative theory, and one can make reliable predictions.

Recently, the general effective field theory of gravity coupled to the standard model of particle physics was constructed~\cite{2019-Ruhdorfer.etal-JHEP}. The authors systematically showed that the first gravity operators appear at mass dimension $6$ in the series expansion, and these operators only couple to the standard model Bosons. They also showed that (i) no new gravity operators appear at mass dimension $7$, (ii) in mass dimension $8$ the standard model Fermions appear, and (iii) coupling between the scalar (Higgs) field and the standard model gauge Bosons appear \emph{only} at mass dimension $8$. 

In the next section, we will consider some specific {MG} models in {4D} and list their salient features. In Sec. \eqref{sec:What?}, we discuss their observational consequences in the upcoming missions.

\section{Modified gravity theories: Examples}\label{sec5}

Earlier in Fig. \eqref{fig:mgoption} we classified {MG} theories according to Lovelock's theorem~\cite{1972-Lovelock-JMP}. As mentioned earlier, in this review, we only consider modifications arising in {4D} space-time. Specifically, we consider two classes of theories --- higher-order derivative theories ($f(R)$, Stelle gravity, Chern-Simons {(CS)} gravity) and scalar-tensor theories. 

\subsection{Classical route to higher derivative gravity}

Assume that we wish to construct a geometrical theory of gravity using the concept of least action. One approach to accomplish this is using Einstein's theory as a paradigm is to write the gravitational action as:
\begin{equation}
S_{\rm Grav} = 
\int \, d^4x \, \sqrt{-g} \, {\cal G}
\label{eq:Gravact}
\end{equation}
where ${\cal G}$ is a scalar that is geometry-dependent or, in other words, is a function of $g_{\mu\nu}$ and its derivatives but is otherwise arbitrary. To keep the discussion as general as possible, we demand that the above action is invariant under arbitrary continuous coordinate transformations:
$$
{x^{\mu}} \to x^{\mu}+\xi^{\mu}
$$
Under the infinitesimal transformation, the action becomes:
$$
\delta S_{\rm Grav} = -2 \int d^{4} x \, \sqrt{-g} \, \xi_{\mu} \,  \nabla_{\nu} {\cal G}^{\mu \nu}
$$
where
\begin{equation}
{\cal G}_{\mu \nu}:=\frac{1}{\sqrt{-g}}\left\{\frac{\partial(\sqrt{-g} {\cal G})}{\partial g^{\mu \nu}}-\partial_{\alpha} \frac{\partial(\sqrt{-g} {\cal G})}{\partial \partial_{\alpha} g^{\mu \nu}}+\partial_{\alpha} \partial_{\beta} \frac{\partial(\sqrt{-g} {\cal G})}{\partial \partial_{\alpha} \partial_{\beta} g^{\mu \nu}}-\ldots\right\}
\label{eq:GravEqu}
\end{equation}
For the action to be stationary under these infinitesimal transformations and $\xi^{\mu}$ being arbitrary, leads to:
\begin{equation}
\nabla_{\nu} {\cal G}^{\mu \nu}=0
\label{eq:Bianchiidentity}
\end{equation}    
In the case of GR, the above expression is contracted Bianchi identities. Since the above expression is general, we can say that the above expression is a consequence of the invariance of the action \eqref{eq:Gravact} under general coordinate transformations~\cite{2016-Tian-GRG}. 

Note that the above expression is generic and does not require us to assume any form of $G$. The two simplest choices of $G$ are (cosmological) constant and $R$ (the Ricci scalar). This is one of the possible scalar invariants, and a plethora of invariants can be constructed for an arbitrary {4D} space-time. If we limit to those invariants whose curvature tensor is quadratic in its ordinary contractions, we then have:
$$
R^{2}, {R}_{\mu \nu} R^{\mu \nu}, \quad R^{\alpha \beta \gamma \delta} {R}_{\alpha \beta \gamma \delta}
$$
Thus, the general higher-derivative gravity theory is given by: 
\begin{equation}
{\cal G} = \Lambda + \frac{1}{2 \kappa^2} R + \alpha R^2 - \beta {R}_{\mu \nu} R^{\mu \nu} 
\end{equation}
where $\alpha$ and $\beta$ are arbitrary coupling parameters, and we have used the Gauss-Bonnet theorem
$$
\delta \int \sqrt{-g} d^{4} x\left(R_{\alpha \beta \gamma \delta} R^{\alpha \beta \gamma \delta}-4 R_{\alpha \beta} R^{\alpha \beta}+R^{2}\right) \equiv 0 \, ,
$$
to rewrite Riemann-square in terms of Ricci scalar and tensor. 

\subsection{Stelle gravity}

The most general quadratic (Stelle) gravity action is:
\begin{equation}
S_{\rm Stelle} 
=\int{d^{4}x \, \sqrt{-g} \left( \frac{1}{2\kappa^{2}}R + \beta R^{2} 
     -\alpha R^{\mu\nu}R_{\mu\nu} \right) }
     \label{eq:StelleAction}
\end{equation}
As opposed to GR which has one mode, Stelle gravity has three modes: one graviton ($1/r$ potential) and two massive Yukawa modes (scalar and spin-2) with masses $\frac{1}{\kappa \sqrt{2\alpha}}$ and $\frac{1}{\kappa \sqrt{4(3\beta-\alpha)}}$
 respectively~\cite{1977-Stelle-PRD,1978-Stelle-GRG}. Demanding that the masses are real and the Hamiltonian is bounded from below lead to the conditions $\alpha\geq 0$, {and} $3\beta\geq\alpha $. It has been shown that Stelle Gravity does not contain malicious ghosts~\cite{1984-Tomboulis-PRL,1986-Antoniadis.Tomboulis-PRD}.

Recently, using a {generalized uncertainty principle} (GUP) modeling maximal momentum, it was shown that the GUP modified dynamics of a massless spin-2 field corresponds to Stelle gravity action \eqref{eq:StelleAction} with suitably constrained parameters~\cite{2021-Nenmeli.etal-PLB}. More specifically, for $\alpha=2\beta = \gamma/\kappa^2$ ($\gamma$ is GUP parameter), Stelle gravity is the \emph{minimally modified, metric-only} theory of gravity which models the effects of maximal momentum. Thus, Stelle gravity can be regarded as the classical manifestation of the imposition of a momentum cutoff at the quantum gravity level.

The mapping provided two physical insights about the Stelle parameters $\alpha$ {and} $\beta$: First, when $\alpha=2\beta$, the additional gauge bosons mentioned earlier have \textit{equal} masses. In this sense, GUP {MG} is a \textit{degenerate} theory. %
Second, Stelle parameters are a measure of the momentum cutoff. Using CMB data, the constraints on the Stelle parameters are~\cite{2021-Nenmeli.etal-PLB}: 
\begin{equation}
10^{-28}GeV^{-2}\leq\gamma\leq 10^{-23}GeV^{-2}
\end{equation}

\subsection{$f(R)$ gravity}
Although Stelle gravity is the simplest modification to GR, the equations are complicated, and it is not easy to obtain exact non-trivial BH solutions. Another simple extension of GR can be obtained by replacing $R$ in the Einstein-Hilbert action \eqref{eq:EHaction} with a general function of $R$, leading to $f(R)$ theories of gravity. The $f(R)$ gravity action is:
\begin{equation}
S_{\rm J} = \frac{1}{2\kappa^2}\int d^4x \, \sqrt{-g} \, f(R) + \int d^4x \sqrt{-g} {\cal L}_m \, .
\label{eq:fRact-Jordan}
\end{equation}
The variation of the above action leads to the following field equation:
\begin{equation}
\label{eq:frfeq}
F R_{\mu \nu}-\frac{1}{2} f(R) g_{\mu \nu}-\nabla_{\mu} \nabla_{\nu} F+g_{\mu \nu} \square F=\kappa^2 T^{(M)}_{\mu \nu}
\end{equation}
where $F \equiv \partial f / \partial R$.
As mentioned above, Bianchi identities ensure that $\nabla^{\mu}T^{(M)}_{\mu \nu}=0$. The trace of the above field equation leads to:
\begin{equation}
\label{eq:fr_Trace}
R F(R)+3 \square F(R)-2 f(R)=\kappa^2 T^{(M)}.
\end{equation}
Since $\square F(R)$ does not vanish for arbitrary $f(R)$, the above equation indicates a propagating scalar degree of freedom, which is absent in GR. 
The following two features make the $f(R)$ gravity an attractive candidate for the alternatives to GR:
\begin{enumerate}
\item $f(R)$ are general enough that higher-order Ricci scalar terms can
encapsulate high energy modifications to GR. Nevertheless, the equations of motion are simple enough that it is possible to solve them! 
\item $f(R)$ theories do not suffer from Ostr\"ogradsky instability~\cite{2007-Woodard-Lect.NotesPhys.}.
\end{enumerate}
The above field equations \eqref{eq:frfeq} contain 
higher-derivative terms, and hence, like Stelle gravity, it is not possible to obtain exact solutions. However, by conformal rescaling of the metric, it 
is possible to transform the above action \eqref{eq:fRact-Jordan} to Einstein-Hilbert action with a minimally coupled scalar field. To avoid instabilities and ghosts, any viable $f(R)$ theory of gravity should satisfy the following conditions~\cite{2007-Starobinsky-JETPLett.}:
\begin{equation}
    \partial f / \partial R > 0, \quad \partial^2 f /\partial R^2 > 0, \quad \mbox{for} \quad R \geq R_0 >0
\end{equation}
where $R_0$ is the present value of $R$.
The first successful $f(R)$ gravity model was the inflationary model proposed by Starobinsky~\cite{1980-Starobinsky-PLB}:
\begin{equation}
f(R)=R+\frac{R^2}{6 M^2}
\end{equation}
where $M$ is a new parameter. This model has two key features:
\begin{enumerate}
\item When $R^2$ dominates, leads to exact de Sitter expansion {without extra scalar field}!
\item When $R$ dominates leads to exit from inflation. 
\end{enumerate}
An approximate solution for the model is given by
  \begin{equation}
  H\simeq H_i -\frac{M^2}{6}(t-t_i)\, ; \quad \quad
  a\simeq a_i e^{[H_i(t-t_i)-(M^2/12)(t-t_i)^2]}.
  \end{equation}
 where $H_i$ is the Hubble parameter and $a_i$  is the scale factor at the beginning of inflation. It has been shown that the Starobinsky model is consistent with the Planck-2018 data~\cite{2020-Akrami.others-Astron.Astrophys.}.
   
   Viable $f(R)$ gravity models have been constructed in the low energy scales. The following models describe the late time acceleration of the Universe ~\cite{2007-Hu.Sawicki-Phys.Rev.D,2007-Starobinsky-JETPLett.}:
\begin{eqnarray}
f(R)&=&R-\mu R_{c} \frac{\left(R / R_{c}\right)^{2 n}}{\left(R / R_{c}\right)^{2 n}+1} \quad n, \mu, R_{c}>0 \\ f(R)&=&R-\mu R_{c}\left[1-\left(1+R^{2} / R_{c}^{2}\right)^{-n}\right] \quad n, \mu, R_{c}>0 .
\end{eqnarray}
Interestingly, the Birkhoff theorem is not valid in $f(R)$ theories~\cite{2020-Xavier.etal-CQG}. Recently, it was shown that an infinite number of exact static spherically symmetric vacuum solutions exist for a class of $f(R)$ gravity. For example, in GR, the zero-spin ($J \to 0$) limit of Kerr {BH} uniquely leads to the Schwarzschild solution. Thus, if there exists a large number of spherically symmetric vacuum solutions in $f(R)$, the results suggest that the no-hair theorem also \emph{may not} hold for $f(R)$ theories. As we discuss in Sec. \eqref{sec:What?}, this has important implications for the future gravitational wave detectors.

{Camanho et al.~\cite{2016-Camanho.etal-JHEP} present one of the most stringent theoretical limits on MG by considering higher derivative corrections to the graviton three-point coupling in a weakly coupled theory of gravity. They argue that the causality constrains these corrections. Additionally, they propose a thought experiment involving a high-energy scattering mechanism that violates causality in terms of the existence of closed time-like curves in these theories. Due to causality conditions, any limited number of terms on the gravitational action functional is ruled out, and it can only be corrected by adding an endless tower of extra massive particles with higher spins~\cite{2017-Edelstein.etal-Phys.Rev.D,2021-Edelstein.etal-JHEP}. Recently, in Refs.~\cite{2020-Chowdhury.etal-JHEP,2021-Chandorkar.etal-JHEP}, the authors use  Classical Regge Growth Conjecture to constrain the extensions to GR. Their objective is to constrain classical theories by establishing a limit on the Regge development of classical scattering amplitudes.  In the context of holography, the so-called chaotic bounds motivate this constraint.}

\subsection{Chern-Simons modified gravity}

GR can be extended in an explicitly parity-violating manner by adding a dynamical pseudoscalar field $\Theta$ and making it couple non-trivially with the curvature. {CS gravity} is a {4D} extension of GR that captures leading-order gravitational parity violation arising from the Pontryagin density~${}^{*}R\, R$~\cite{Jackiw:2003pm,Alexander:2009tp}. Thus, the action is 
\begin{equation}
\mathcal{S}=\int d^4x \sqrt{-g} \left[\frac{R}{2\kappa^2} + \frac{\alpha}{4} \Theta {}^*RR - \frac{\beta}{2} \left(\nabla\Theta\right)^2\right]; 
\quad\mbox{where}\quad
 {}^*RR = \frac{1}{2}R_{\mu\nu\rho\sigma}\epsilon^{\mu\nu\alpha\beta}R_{\alpha\beta}^{\quad\rho\sigma} \, .
\end{equation} 
 CS theories are of two types: First is referred to as {canonical} CS~\cite{Jackiw:2003pm}.
In this case, $ \Theta $ is a constant with no kinetic and potential term. The second is referred to as dynamical CS (dCS). In this case,  $ \Theta $ is a fully dynamical field~\cite{Smith:2007jm}. 
Parity violating theories introduce corrections to the frame-dragging effects of spinning bodies and hence have been used as an explanation for the anomalous gyroscopic precession observed in the LAGEOS mission~\cite{Smith:2007jm}.

For spherically symmetric space-times, the Pontryagin density vanishes, leading to standard GR with a scalar field potential. Owing to the no-hair theorem, for a spherically symmetric background, the only stable solution possible is the Schwarzschild. Hence, Schwarzschild space-time is a solution of both kinds of CS theories \cite{Jackiw:2003pm}. In the case of axisymmetric solutions, the Pontryagin density does not vanish, and hence, axisymmetric solutions are non-trivial to construct in CS theories~\cite{Delsate:2018ome}. So far in the literature, no closed-form fast-spinning Kerr-like {analytic} solution exists in either kind of CS theory. However, it is possible to construct axisymmetric solutions from spherically symmetric solutions perturbatively in spin~\cite{Yunes:2009hc}.

%
%
Like Schwarzschild, FLRW space-time \eqref{eq:FRW} is an exact solution. This is because Pontryagin density vanishes for homogenous and isotropic space-times. Interestingly, {CS} modification can provide a mechanism for Baryogenesis~\cite{2006-Alexander.etal-PRL}.
In Sec. \eqref{sec:What?}, we show that it is possible to distinguish CS and GR using the QNM frequencies of the resultant BH from binary mergers. 

\subsection{Scalar-tensor gravity theories}

Among {MG} theories, scalar-tensor theories have received more attention than others. 
Although the scalar fields are hypothetical, they naturally occur in the standard model of particle physics and unified field theories starting from Kaluza-Klein. Jordan's
starting point was Kaluza-Klein's unified field theory in a \red{5D} space-time which can be rewritten as projective {4D} space-time with extra space dimension appearing as a space-time scalar. ({for a historical review}, see \cite{2005-Brans-Talk}.] Interestingly, this scalar field also nicely fitted with Dirac's Large number hypothesis ~\cite{1937-Dirac-nat,1938-Dirac-PRSA,2005-Brans-Talk}. Dirac was intrigued by \emph{coincidental approximate} equality between important physical quantities expressed in a dimension-free manner, i. e.,
\begin{eqnarray}
& & \frac{\mbox{electrical force between electron and proton}}{\mbox{gravitational force between electron and proton}} = \frac{e^2}{G m_p m_e}  \approx 10^{40} \\[10pt]
& & \frac{\mbox{Age of the Universe}}{\mbox{time taken by light to cross an atom}} = \frac{t_{\rm Univ}}{e^2/(m_e \, c^2)} \approx 10^{40}
\end{eqnarray}
Assuming that these {hold} at all times after the instant of the Big Bang and the atomic parameters do not vary with time, Dirac went on to suggest that 
\begin{equation}
\frac{e^2}{G m_p m_e}  \propto t_{\rm Univ} \quad \Longrightarrow \quad 
G \propto \frac{1}{t_{\rm Univ}}
\end{equation}
Since the mass of elementary particles and the charge of the electron is fixed to be constants by experiments, the above relation indicates that $G$ can vary in time. This possibility was explored in the scalar-tensor theory: Jordan-Brans-Dicke theory, which introduced a time-varying gravitational ``constant" represented by a scalar field $\varphi$~\cite{1961-Brans.Dicke-Phys.Rev.}. This is the simplest way of introducing a time-dependent gravitational constant.
\begin{equation}
S_{JBD} =\frac {1}{16\pi} \int d^{4}x \, \sqrt{-g} 
\left( \varphi R- \frac{\omega_{\rm BD} }{\varphi} \partial_{\mu}\varphi \partial^{\mu}\varphi \right)+\int d^{4}x \sqrt {-g} L_{\mathrm {M}}
\end{equation}
where $\omega_{BD}$ is the dimensionless Brans-Dicke coupling constant.
Gravitational dynamics in the Jordan-Brans-Dicke theory is dictated by the following equations:
\begin{equation}
\small{
\begin{aligned}
& R_{\mu \nu}(g)-\frac{1}{2} g_{\mu \nu} R(g)= 8\pi\frac{T_{\mu \nu}}{\varphi} 
+\frac{1}{\varphi}\left(\nabla_{\mu} \nabla_{\nu} \varphi-g_{\mu \nu} \square \varphi\right) 
+\frac{\omega_{\mathrm{BD}}}{\varphi^{2}}\left[\partial_{\mu} \varphi \partial_{\nu} \varphi-\frac{1}{2} g_{\mu \nu}(\nabla \varphi)^{2}\right] \\
& \left(3+2 \omega_{\mathrm{BD}}\right) \square \varphi= 8 \pi T
\end{aligned}
}
\end{equation}
Through the variation of $\varphi$, the relation $G \propto t_{\rm Univ}^{-1}$ can be realized . 

Scalar tensor theories also describe the early Universe inflationary phase in which a scalar field (inflaton) drives the inflationary expansion~\cite{2000-liddle.lyth-Book,2007-Lemoine.etal-Book,2015-Amendola.Tsujikawa-DarkEnergyTheory}.
With the discovery of the accelerated expansion of the late Universe, there has been renewed interest in the scalar-tensor theories of gravity. They were proposed to generalize the $\Lambda$CDM model, with the scalar field representing the dynamic dark energy component that drives the accelerated expansion.

A variety of scalar field dark energy models discussed are proposed in the literature, like 
canonical scalar field (also known as quintessence)~\cite{1988-Ratra.Peebles-Phys.Rev.D}, kinetic dominated scalar fields (k-essence)~\cite{1999-Armendariz-Picon.etal-PLB}, Galileons~\cite{2009-Nicolis.etal-Phys.Rev.D}. The simplest among these is \textit{quintessence} whose action is:
\begin{equation}
\label{eq:quint_action}
S = \int d^4x \sqrt{-g}\left(\dfrac{1}{2 \kappa^2}R-\dfrac{1}{2}g^{\mu \nu}\nabla_{\mu}\phi \nabla_{\nu}\phi - U(\phi) \right)+S_{m}.
\end{equation}
One can define the background energy density and pressure of the scalar field as
\begin{equation}
\label{eq:phi_rho_p}
\rho_{\phi}=\dfrac{1}{2}\dot{\phi}^2+U(\phi), \quad p_{\phi}=\dfrac{1}{2}\dot{\phi}^2-U(\phi)
\end{equation}
leading to the time-varying equation of state
\begin{equation}
\label{eq:wphi}
w_{\phi}=\dfrac{\dot{\phi}^2-2U(\phi)}{\dot{\phi}^2+2U(\phi)}
\end{equation}
From the above expression, we see that the $w_{\phi}$ can take the values in the range $-1 \leq w_{\phi} \leq 1$. In case of a near flat potential, for which $\dot{\phi}\simeq 0$, quintessence mimics the cosmological constant with $w_{\phi}\simeq-1$.
The evolution of the scalar field is determined by the equation of motion
\begin{equation}
\label{eq:eom_phi}
\ddot{\phi}+3H\dot{\phi}+U_{\phi}(\phi)=0 \quad \mbox{or} \quad \dot{\rho}_{\phi} + 3H(1+w_{\phi})=0
\end{equation}
The evolution in a quintessence model is determined by the form of the scalar field potential. For example, in quintessence models which satisfy the condition $\Gamma \equiv UU_{\phi\phi}/U_{\phi}^2 \geq 1$, the scalar field follows a common evolutionary path for a wide range of initial conditions~\cite{2015-Amendola.Tsujikawa-DarkEnergyTheory}. Hence these models, called `tracker models,' alleviate the coincidence problem. Examples of the tracker potential, which leads to the accelerated expansion, include the modified exponential potential $U(\phi) =U_0 [\cosh(\lambda \phi)-1]^p$, and the inverse power-law potential given by $U(\phi) = U_0 \phi^{-\alpha}$~\cite{1988-Ratra.Peebles-Phys.Rev.D}. In this class of models, initially, the field rolls down the potential and slows down after the Universe enters the phase of accelerated expansion.

The {scalar} tensor theory action \eqref{eq:quint_action} is related to $f(R)$ theory action \eqref{eq:fRact-Jordan} by the mapping:
\begin{equation}
\varphi = F(R); \quad U(\varphi) = \frac{R \, F - f}{2} \quad \Longrightarrow \quad 
\omega_{BD} = 0
\end{equation}
It is consistent with the fact that $f(R)$ gravity theories have an extra scalar degree of freedom.
 
\subsection{Extending scalar-tensor theory}

We can extend the minimally coupled canonical scalar-tensor theories of gravity by modifying various terms of the scalar-tensor action and/or introducing new terms. A few examples of such theories are given below.
\begin{itemize}
\item Generalized Kinetic scalar field (e.g., K-essence) 
$$\mathcal{L}_{\rm Can} =X-V(\phi) \quad \Longrightarrow \quad 
\mathcal{L}_{\rm NC} =K(\phi, X) \quad \mbox{where} \quad 
X= \frac{1}{2} g^{\mu \nu} \partial_{\mu} \phi \partial_{\nu} \phi 
$$
%
This model was originally proposed as an inflationary model~\cite{1999-ArmendarizPicon.etal-Phys.Lett.B}, later applied to the late time accelerated expansion of the Universe~\cite{2000-Chiba.etal-Phys.Rev.D}. It has been shown that k-essence models do not suffer from the fine-tuning problem that plagues the $\Lambda$CDM model.
\item Generalized non-minimal coupling with gravity 
$$
S=\int d^{4} x \sqrt{-g}\left[ \frac{1}{2 \kappa^2} R+\mathcal{L}_{\phi}\right]  \quad  \Longrightarrow \quad 
S_{\rm NM} =\int d^{4} x \sqrt{-g} \left[ f(\phi) R + \mathcal{L}_{\phi}\right]
$$
Here, there is a non-minimal coupling between the scalar field and the curvature term. Examples of this type of theory include JBD theory and Higgs inflationary model~\cite{2008-Bezrukov.Shaposhnikov-Phys.Lett.B}. Scalar tensor theories with non-minimal coupling can also describe dark energy - dark matter interaction~\cite{1999-Amendola-Phys.Rev.D}.

\item Including higher derivatives (e.g., Galileons)

$$
\mathcal{L}_{\rm NC} =K(\phi, X) \quad  \Longrightarrow \quad  
\mathcal{L}_{\rm Gal}= K(\phi, X) + G(\phi, X) \square \phi
$$
In these models, the action contains $2^{nd}$ or higher-order derivatives of the scalar field. The Galileon model, which was motivated by instabilities in the \red{5D} DGP gravity, is a good illustration of this.~\cite{2009-Nicolis.etal-Phys.Rev.D}. They are constructed in such a way that the equations of motion contain only upto $2^{nd}$ derivative terms. A more general example of the higher derivative theory is the Horndeski theory~\cite{1974-Horndeski-Int.J.Theor.Phys.}. It is constructed to be the most general theory of gravity, with action being a function of metric and scalar field that lead to $2^{nd}$ order equations of motion. The scalar-tensor theories mentioned in this section and GR are special cases of Horndeski theories.
\end{itemize}

It must be noted that theories with higher-order derivatives can result in ghosts. One needs to bypass the Ostr\"ogradsky instability to avoid these instabilities~\cite{2007-Woodard-Lect.NotesPhys.}. This is discussed in the next section.

\subsection{By-passing Ostr\"ogradsky instability}
As seen in the previous sections, one way to construct {MG} theories is by introducing higher derivative terms to existing theories. However, these often lead to instabilities. This section discusses avoiding these instabilities by bypassing the Ostr\"ogradsky instability.

First, we check if Einstein-Hilbert action has any ghost modes. The action \eqref{eq:EHaction} can be rewritten in terms of the Riemann tensor as
\begin{equation}
{\cal S}_{\rm EH} = 
\frac{1}{2 \kappa^2} \int d^{4} x \sqrt{-g} R = 
\frac{1}{2 \kappa^2} \int d^{4} x \sqrt{-g} R_{\mu \nu} g^{\mu \nu} = 
\frac{1}{2 \kappa^2} \int d^{4} x \sqrt{-g} \,  R_{\mu \lambda \nu}^{\lambda} \, g^{\mu \nu}.
\end{equation}
Expanding the Riemann tensor, we have
\begin{eqnarray}
\nonumber
R_{\mu \lambda \nu}^{\lambda} &=&\partial_{\lambda} \Gamma_{\mu \nu}^{\lambda}-\partial_{\nu} \Gamma_{\lambda \mu}^{\lambda}+ 
\mbox{$1^{st}$ order derivatives} \\
\nonumber
&=&\frac{1}{2} \partial_{\lambda}\left(g^{\lambda \rho}\left(\partial_{\mu} g_{\nu \rho}+\partial_{\nu} g_{\rho \mu}-\partial_{\rho} g_{\mu \nu}\right)\right)-\frac{1}{2} \partial_{\nu}\left(g^{\lambda \rho}\left(\partial_{\mu} g_{\lambda \rho}\right)\right)+ \text{$1^{st}$ order derivatives}  \\
\nonumber
&=&\frac{1}{2} g^{\lambda \rho}\left(\partial_{\lambda} \partial_{\mu} g_{\nu \rho}+\partial_{\lambda} \partial_{\nu} g_{\rho \mu}-\partial_{\lambda} \partial_{\rho} g_{\mu \nu}-\partial_{\mu} \partial_{\nu} g_{\lambda \rho}\right)+
\text{{$1^{st}$ order derivatives}} 
\end{eqnarray}

Substituting this back in action, we get:
\begin{equation}
{\cal S}_{\rm EH} = \frac{1}{2 \kappa^2} \int d^{4} x \sqrt{-g} \frac{1}{2} g^{\mu \nu} g^{\lambda \rho}\left(\partial_{\lambda} \partial_{\mu} g_{\nu \rho}+\partial_{\lambda} \partial_{\nu} g_{\rho \mu}-\partial_{\lambda} \partial_{\rho} g_{\mu \nu}-\partial_{\mu} \partial_{\nu} g_{\lambda \rho}\right)+
\mbox{{$1^{st}$ order derivatives}}
\end{equation}
Here the $2^{nd}$ order derivative terms can be cast into terms containing only the first derivative using the product rule. The first term above can be recast:
\begin{equation}
\int d^{4} x \sqrt{-g} \frac{1}{2} g^{\mu \nu} g^{\lambda \rho} \partial_{\lambda} \partial_{\mu} g_{\nu \rho}=\int d^{4} x \partial_{\lambda}\left[\sqrt{-g} \frac{1}{2} g^{\mu \nu} g^{\lambda \rho} \partial_{\mu} g_{\nu \rho}\right] 
-\int d^{4} x \partial_{\lambda}\left[\sqrt{-g} \frac{1}{2} g^{\mu \nu} g^{\lambda \rho}\right] \partial_{\mu} g_{\nu \rho}
\end{equation}
Here we see that the first term is a total divergence term, and the second term contains only the first derivatives of the metric. Thus Einstein-Hilbert action does not have ghost modes.

Next, we look at how to bypass the Ostr\"ogradsky instability for a generic model. One of the key requirements for the Ostr\"ogradsky theorem to be applicable is the non-degeneracy. The non-degeneracy condition refers to $\partial L / \partial \ddot{q}$ to be dependent on $\ddot{q}$. One way to break the degeneracy is to consider the Lagrangian for which $\partial^2 L / \partial\ddot{q}^2 = 0$. In case Lagrangian depends on only a position $q$ and its velocity $\dot{q}$, then the equations of motion are first order, which represents not the dynamics but the constraint. If Lagrangian depends on $q, \dot{q}, \ddot{q}$,
degeneracy implies that equations of motion will have second-order or higher derivatives, representing the dynamics~\cite{2007-Woodard-Lect.NotesPhys.}. 

Example of these type of models include Galileons which are hypothetical scalar fields whose Lagrangians involve multilinear terms of first and
second derivatives, but nonlinear field equations are second order. A simple example with a single scalar field $\pi$ is given by
\begin{equation}
S_{G}=\int d^{n} x \, \partial_{\alpha}\pi \, \partial^{\alpha}\pi \, 
\partial_{\beta} \partial^{\beta} \pi 
\end{equation}
Here $S_G$ is invariant under the transformation $\pi\, \rightarrow\, \pi\, +\, b_{\mu}x^{\mu}\, +\, c$ where $b_\mu$ is a constant vector. Note that this is similar to the Galilean transformation $\dot{\vec{x}} \rightarrow \dot{\vec{x}} + \vec{V}$ in particle mechanics.

The action $S_G$ {leads} to the field equation
\begin{equation}
(\partial_{\beta} \partial^{\beta} \pi) \, ( \partial_{\alpha} \partial^{\alpha} \pi)
- (\partial_{\alpha} \partial_{\beta} \pi) \, 
(\partial^{\alpha} \partial_{\alpha} \pi) = 0,
\end{equation} 
which is invariant under the transformation $c+b_{\mu} x_{\mu}$.

One can have  Galileon models with non-canonical and non-minimally coupled single-field models leading to second-order equations of motion:
\begin{eqnarray}
\nonumber
\mathcal{L}_{2}&=& K(\pi, X), \qquad \mbox{where} 
\quad X=\frac{1}{2}(\nabla \pi)^{2}, \quad G_{i X} \equiv \partial G_{i} / \partial X 
\\ 
\nonumber
\mathcal{L}_{3}&=&-G_{3}(\pi, X) \square \pi \\
\nonumber 
\mathcal{L}_{4}&=& G_{4}(\pi, X) R+G_{4 X}\left[(\square \pi)^{2} 
-\left(\nabla_{\mu} \nabla_{\nu} \pi\right)^{2}\right] \\ 
\nonumber
\mathcal{L}_{5}&=& G_{5}(\pi, X) G_{\mu \nu} \nabla^{\mu} \nabla^{\nu} \pi - 
\frac{1}{6} G_{5 X}\left[(\square \pi)^{3}-3(\square \pi)\left(\nabla_{\mu} \nabla_{\nu} \pi\right)^{2}+2\left(\nabla_{\mu} \nabla_{\nu} \pi\right)^{3}
\right] . 
\end{eqnarray} 
Many of inflation and dark energy models can be understood in a unified manner. For example, $ G_4 = 1/(2 \kappa^2)$ yields the Einstein-Hilbert action;  $G_4 = f(\pi)$ yields a non-minimal coupling ; $G_5 \propto \pi$ corresponds to new Higgs inflation.

Horndeski gave the most general {4D} action constructed from the metric 
$\mathrm{g}$, the scalar field $\phi$, and their derivatives, $\partial g_{\mu \nu}, \partial^{2} g_{\mu \nu}, \partial^{3} g_{\mu \nu}, \cdots, \partial \phi, \partial^{2} \phi, \partial^{3} \phi, \cdots$ leading to $2^{nd}$ order equations:
\begin{eqnarray}
\mathcal{L}_{H} &=& \delta_{\mu \nu}^{\alpha \beta \gamma}\left[\kappa_{1} \nabla^{\mu} \nabla \alpha \phi R_{\beta \gamma}{ }^{\nu \sigma}+\frac{2}{3} \kappa_{1 X} \nabla^{\mu} \nabla \alpha \phi \nabla^{\nu} \nabla_{\beta} \phi \nabla^{\sigma} \nabla_{\gamma} \phi+\kappa_{3} \nabla_{\alpha} \phi \nabla^{\mu} \phi R_{\beta \gamma}{ }^{\nu \sigma} \right] \nonumber \\
&+&  \left. 2 \kappa_{3 X} \nabla_{\alpha} \phi \nabla^{\mu} \phi \nabla^{\nu} \nabla_{\beta} \phi \nabla^{\sigma} \nabla_{\gamma} \phi\right] 
- 6\left(F_{\phi}+2 W_{\phi}-X \kappa_{8}\right) \square \phi+\kappa_{9} 
\nonumber \\
&+& \delta_{\mu \nu}^{\alpha \beta}\left[(F+2 W) R_{\alpha \beta}{ }^{\mu \nu}+2 F_{X} \nabla^{\mu} \nabla_{\alpha} \phi \nabla^{\nu} \nabla_{\beta} \phi+2 \kappa_{8} \nabla_{\alpha} \phi \nabla^{\mu} \phi \nabla^{\nu} \nabla_{\beta} \phi\right] 
\end{eqnarray}
where $\kappa_1, \kappa_3, \kappa_8, \kappa_9, {F}$ are functions of $\phi, X$. 
$F_{X}=2\left(\kappa_{3}+2 X \kappa_{3 X}-\kappa_{1 \phi}\right)$, $W = W(\phi)$.
However, certain Hordenski models have been ruled out after BNS merger observations~\cite{2017-Sakstein.Jain-PRL,2017-Baker.etal-PRL,2018-Crisostomi.Koyama-PRD}.

\section{What are the observational consequences?}
\label{sec:What?}

From the above discussions, it is clear that there are many ways of modifying GR in the strong-gravity and cosmological distances (such as Stelle gravity, $f(R)$, scalar-tensor and {CS} gravity), and each model has its unique features. This leads to the following crucial question: Is there a set of unique signatures that {distinguish {MG} theories from GR}? This section investigates this issue by considering three specific cases: Testing {MG} theories using gravitational wave (GW) observations~\cite{2017-Bhattacharyya.Shanki-PRD,2018-Bhattacharyya.Shanki-EPJC,2019-Bhattacharyya.Shanki-PRD,2019-Shankaranarayanan-IJMPD,2021-Srivastava.etal-PRD}, 
{low-energy} observational signatures in cosmological observations~\cite{2019-Johnson.Shankaranarayanan-Phys.Rev.D} and {non-minimal} coupling of electromagnetic fields leading to observable signatures~\cite{2020-Kushwaha.Shanki-PRD,2021-Kushwaha.Shanki-PRD,2022-Jana.Shanki}. 

\subsection{Testing modified gravity theories in strong gravity regimes}

According to GR, three measurable quantities (mass, charge, and angular momentum) fully characterize the isolated BHs in equilibrium~\cite{1998-Bekenstein-Talk}. In other words, any \emph{deformations of the {horizon}} will {finally settle into a BH configuration that is described \emph{only} by the above three quantities.}~\cite{1998-Bekenstein-Talk}.
When two BHs merge to form another BH, the remnant BH's event horizon {will be} highly distorted. As a result, it radiates GWs until it settles into an equilibrium configuration~\cite{2009-Sathyaprakash.Schutz-LRR,2013-Yunes.Siemens-LRR}. GWs emitted, referred to as quasi-normal modes (QNMs), are a superposition of damped sinusoids and depend only on the parameters characterizing the BH, namely, {its} mass and spin (astrophysical BHs are not likely to be electrically charged)~\cite{1999-Nollert-CQG,1999-Kokkotas.Schmidt-LRR,2011-Konoplya.Zhindeko-RPP}. 
QNMs are the fingerprints of the final BH. The simplicity of the {QNM} spectrum allows one to identify the Kerr solution~\cite{2015-Berti.etal-CQG}. If GR is still valid at strong gravity regimes, then the final BH will be consistent with the no-hair theorem. However, this will not be the case {for the {MG} theories} if GR is modified. Broadly, there are two approaches in which we can test strong field limit of gravity: 
\begin{enumerate}
    \item Obtain new BH solutions for the {MG} models. Use the template matching technique to match GW signals and identify the deviations. 
    \item Identify {MG} theories with the same BH solutions as in GR. Obtain the difference in GW signals in GR and {MG} theories. 
\end{enumerate}
Recently, it has been explicitly shown that the Birkhoff theorem is not valid in $f(R)$ theories~\cite{2020-Xavier.etal-CQG}. Thus, if there exists a large number of spherically symmetric vacuum solutions in $f(R)$, the results suggest that the no-hair theorem also \emph{may not} hold for $f(R)$ theories. However, we need to test this conjecture for a larger class of {MG} models. 

{As mentioned earlier, it is known that Kerr is not a solution for the CS theory. Besides, no closed-form fast-spinning Kerr-like {analytic} solution exists in either kind of CS theory. However, it is possible to construct axisymmetric solutions from spherically symmetric solutions perturbatively in spin~\cite{Yunes:2009hc}. Unlike in $f(R)$, even in the slow-rotating limit, the rotating solution in CS is different from Kerr. In 
other words, no hair theorem may not be a generic property for MG models. There have been efforts in the literature to test the no hair theorem using the first and higher overtones in GW observations~(see, for instance, 
Refs.~\cite{2019-Isi.etal-Phys.Rev.Lett.,2020-Islam.etal-PRD}). Recently, 
Psaltis et al. used numerical calculations of shadows and showed that
spacetimes that deviate from the Kerr metric could lead to large changes in the predicted shadows~\cite{2020-Psaltis.others-Phys.Rev.Lett.}.}

The spectrum of QNM predicted by GR comprises of two isospectral towers of modes that are respectively even and odd under parity~\cite{1985-Chandrasekhar-Book}. Also, GWs emit equal energy in the two polarization states~\cite{1985-Chandrasekhar-Book}. As seen before, for modified theories of gravity, the total emitted energy gets redistributed due to additional propagating degrees of freedom, and the two polarization states need not carry equal energy. Besides extra polarization modes, alternative gravity theories can introduce three different effects on QNMs~\cite{2019-Barack.etal-CQG}:
\begin{enumerate}
\item modify the spectrum of even and odd modes while preserving isospectrality.
\item break isospectrality (do not emit GWs with equal energy in the two polarization states) 
\item mix the even and odd modes so that there is no longer a distinction between the two. 
\end{enumerate}
In the rest of this section, we heuristically show that polarization provides a unique tool to distinguish between GR and modified theories of gravity. Consider a spherically symmetric space-time:
\begin{equation}
ds^2 = - g(r) dt^2 + \frac{dr^2}{h(r)} + r^2 \left(d\theta^2 + \sin^2 \theta 
d\phi^2 \right) \, ,
\label{eq:SpSyST}
\end{equation} 
where $g(r)$ and $h(r)$ are arbitrary functions of $r$. At the linear order, the two modes of metric perturbations ($\Phi_1$ and $\Phi_2$) of the above space-time satisfy the following equations~\cite{1985-Chandrasekhar-Book}:
\begin{eqnarray}
\frac{d^2\Phi_i}{dr_*^2} + \left[ \omega^2-V_i(r_*)  \right] \Phi_i = 0 ~~~ & \mbox{where} & ~~~
i= 1, 2 \, , \label{eq:GRper}
\end{eqnarray}
$r_* = \int dr/\sqrt{h(r) g(r)}$ is referred to as \emph{tortoise coordinate}, $\omega$ are the complex frequencies of the quasi-normal modes, and $V_i(r_*)$ are the potentials corresponding to the two modes ($\Phi_i$). The two potentials are related to each other via Darboux transformations~\cite{1985-Chandrasekhar-Book}. Hence,  the modes have the same reflection coefficients and carry equal energy~\cite{1985-Chandrasekhar-Book}. 

As discussed in Sec.~\eqref{sec4}, the equations of motion of any generic {MG} can be written in the following compact form: 
\begin{equation}
{\cal G}_{\mu\nu} = \kappa^2 \, T_{\mu\nu} \, ,
\end{equation} 
where ${\cal G}_{\mu\nu}$ is the modified Einstein tensor.  
Following Eq. \eqref{eq:Bianchiidentity}, in {MG}, 
the energy-momentum tensor is locally conserved ($\nabla^{\mu} T_{\mu\nu}  = 0$). For $f(R)$ gravity, the generalized Bianchi identity \eqref{eq:Bianchiidentity} leads to:
\begin{equation} 
\label{eq:BianchifR}
f^{\prime\prime}(R) \left( R_{\mu\nu} \nabla^{\mu} R \right) = 0 \, . 
\end{equation}
For GR, $f(R) = R$. Hence, $f^{\prime\prime}(R)$ vanishes and the above equation is trivially satisfied. However, $f^{\prime\prime}(R)$ is non-zero 
for {MG} theories, hence, the generalized Bianchi identity (\ref{eq:BianchifR}) leads to four constraints on the Ricci tensor. While, GR and $f(R)$ have four constraints on the field variables, the number of dynamical variables are different. For $f(R)$ gravity, unlike GR, the trace equation \eqref{eq:fr_Trace} is dynamical. 

As a result, $f(R)$ gravity has 11 dynamical variables --- 10 metric variables ($g_{\mu\nu}$) and Ricci scalar ($R$). However, GR has only ten metric variables ($g_{\mu\nu}$). In other words, in $f(R)$, the scalar curvature $R$ plays a non-trivial role in determining the metric itself. Since this extra degree of freedom is a scalar, it can be treated as longitudinal mode~\cite{2017-Bhattacharyya.Shanki-PRD}.  

Interestingly, the extra propagation mode is a generic feature for any pure curvature-modified theories of gravity containing only the higher-order Ricci scalar/tensor terms and without any additional matter fields. The generalized Bianchi identity (\ref{eq:Bianchiidentity})  leads to four non-trivial constraints between the Ricci tensor and Ricci scalar. Thus, any modifications to GR will have at least \emph{one extra dynamical field} that plays a non-trivial role in the determination of the metric itself.

The crucial point in the case of {MG} theories is that as remnant BH settles down to an equilibrium state, some energy will be carried by the extra dynamical fields (say, $\Psi_{\rm new}$). This \emph{missing energy} signals modifications to gravity. This leads us to the crucial question: How the missing energy can be used to distinguish GR and {MG} theories using gravitational wave detectors? 

As mentioned above, in the case of GR, the two modes of perturbations satisfy Eq. (\ref{eq:GRper}). However, for {MG} theories, the two modes of perturbations for spherically symmetric space-times (\ref{eq:SpSyST}) --- that is related to the two polarizations detected by gravitational-wave detectors --- satisfy the following relations~\cite{2017-Bhattacharyya.Shanki-PRD,2018-Bhattacharyya.Shanki-EPJC,2019-Bhattacharyya.Shanki-PRD}:
\begin{eqnarray}
\frac{d^2\Phi_i}{dr_*^2} + \left(\omega^2-V_i\right)\Phi_i = 
S^{eff}_i\left(\Psi_{\rm new}\right) & & \qquad i= 1, 2 
\label{eq:MGper}
\end{eqnarray}
where $S^{eff}_i(\Psi_{\rm new})$ are the effective source terms comprising of the new degrees of freedom. In general, $S^{eff}_1(\Psi_{\rm new}) \neq S^{eff}_2(\Psi_{\rm new})$.  For $f(R)$ theories it can be shown that $S^{eff}_2(\Psi_{\rm new})$ vanishes while $S^{eff}_2(\Psi_{\rm new})$ is non-zero~\cite{2017-Bhattacharyya.Shanki-PRD,2018-Bhattacharyya.Shanki-EPJC,2019-Bhattacharyya.Shanki-PRD}. Thus, the two modes of perturbation in {MG} theories \emph{do not} satisfy isospectral relation ({$\Phi_1 \neq \Phi_2$}) leading to an energetic inequality between the two observable modes in gravitational wave detectors. This energy inequality can be parameterized by a \emph{strong gravity diagnostic  parameter}~\cite{2017-Bhattacharyya.Shanki-PRD,2018-Bhattacharyya.Shanki-EPJC,2019-Bhattacharyya.Shanki-PRD}:
\begin{eqnarray}
\Delta &=& \frac{\left|d_{t,\Omega}\Phi_1 \right|^2 - \left|d_{t,\Omega}\Phi_2\right|^2}{\left|d_{t,\Omega}\Phi_1 \right|^2 + \left|d_{t,\Omega}\Phi_2\right|^2} \label{SGdiagnostic} 
\end{eqnarray}
where the $ d_{t,\Omega} $ corresponds to derivative with respect to time and solid angle. {This parameter vanishes only when $\Phi_1 = \Phi_2$,} hence, vanishes \emph{only} for {GR}. For {MG} theories, $\Phi_1 \neq \Phi_2$ and {the parameter} is non-zero.  In the case of $f(R)$ gravity, $\Delta \sim 10^{-7}$ for 
$10 M_{\odot}$ BHs~\cite{2017-Bhattacharyya.Shanki-PRD,2018-Bhattacharyya.Shanki-EPJC}. In the next generation of gravitational-wave detectors (e.g. the Cosmic Explorer \cite{2017-Abbott.etal-CQG}) the signal-to-noise ratio in the quasi-normal mode regime alone could be as large as ${\rm SNR} > 50$~\cite{2015-Nakano.etal-PRD}. With such detectors, the above diagnostic parameter can provide a unique signature for strong gravity. 

In the case of slowly-rotating {BH} solution in {dCS} gravity, it was recently shown that dCS corrections are potentially observable when the final BH mass is less than $15 M_{\odot}$~\cite{2021-Srivastava.etal-PRD}. Thus, it was shown that future binary Neutron Star events could potentially distinguish dCS and GR.

\subsection{Low-energy observational signatures of $f(R)$ gravity}

Many different $f (R)$ models have been proposed to account
for the late-time acceleration of the Universe. Similarly, many different dark energy models within GR can
also account for the late-time acceleration. This leads to the question: Are there signatures distinguishing dark energy and {MG} models? We can try to answer this in two ways. The first method assumes a specific form of $f(R)$ and looks for its signatures in the observational data. Another option is to keep the form of $f(R)$ arbitrary and construct the tools to detect the departure from GR using the observational data. Here we present the latter method, which is model independent~\cite{2019-Johnson.Shankaranarayanan-Phys.Rev.D}.

Using the local distance measurements, model independent expansion history of the late Universe has been constructed~\cite{2018-Shafieloo.etal-Phys.Rev.D}. Any cosmological model that describe the evolution of the late Universe should be consistent with the background expansion history. Hence we cannot use background evolution to distinguish between {MG} theories and GR. However, the evolution of scalar perturbations can be a useful tool for this purpose. Consider the perturbed FLRW space-time in the Newtonian gauge
\begin{equation}
    d s^{2}=-(1+2 \Phi) d t^{2}+a^{2}(t)(1-2 \Psi) \delta_{i j} d x^{i} d x^{j}
\end{equation}
where $\Phi \equiv \Phi\left(t, x^{i}\right)$ and $\Psi \equiv \Psi\left(t, x^{i}\right)$ are the scalar perturbations. Given the background expansion history, the evolution of the function $F(R) \equiv \partial f / \partial R$ can be obtained using the background field equation:
\begin{equation}
     \ddot{F}-H \dot{F}+2 F \dot{H}=-\kappa^2 \left(\rho_{M}+P_{M}\right)
\end{equation}
Bianchi identities ensure that the continuity equation for the matter fluid is applicable to both GR and $f(R)$ gravity
\begin{equation}
    \dot{\bar{\rho}}+3 H (\bar{\rho} + \bar{P})=0
\end{equation}
Even though the background evolution is identical in both scenarios, there are significant differences in the evolution of the scalar perturbations. In GR, the evolution of the perturbations is determined by the equations:
\begin{equation} 
\ddot{\delta}_{\mathrm{GR}}+2 H \dot{\delta}_{\mathrm{GR}}-\frac{\kappa^{2}}{2} \bar{\rho} \delta_{\mathrm{GR}};~~ 
\frac{k^{2}}{a^{2}} \Phi_{\mathrm{GR}}+\frac{\kappa^{2}}{2} \bar{\rho} \delta_{\mathrm{GR}} =0;~~ \Phi_{\mathrm{GR}}-\Psi_{\mathrm{GR}} =0 
\end{equation}
In $f(R)$ gravity, the scalar perturbations satisfy the following equations~\cite{2019-Johnson.Shankaranarayanan-Phys.Rev.D}:
\begin{align} 
& \Phi_{MG}-\Psi_{MG} = -\dfrac{\delta F}{\bar{F}};
\qquad \ddot{\delta}_{\mathrm{MG}}+2 H \dot{\delta}_{\mathrm{MG}}-\frac{\kappa_{\mathrm{eff}}^{2}}{2} \bar{\rho} \delta_{\mathrm{MG}} =0 \nonumber \\
& \dot{\Psi}_{\mathrm{MG}}+\left(H-\frac{F \dot{H}}{\dot{F}}+\frac{F}{3 \dot{F}} \frac{k^{2}}{a^{2}}\right) \Phi_{\mathrm{MG}}+\left(\frac{F \dot{H}}{\dot{F}}+\frac{F}{3 \dot{F}} \frac{k^{2}}{a^{2}}\right) \Psi_{\mathrm{MG}}+\frac{\kappa^{2} \bar{\rho}}{3 \dot{F}} \delta_{\mathrm{MG}} =0 \\
& \dot{\Phi}_{\mathrm{MG}}+\left(H-\frac{\dot{F}}{F}-\frac{F \dot{H}}{\dot{F}}-\frac{F}{3 \dot{F}} \frac{k^{2}}{a^{2}}\right) \Psi_{\mathrm{MG}}+\left(2 \frac{\dot{F}}{F}+\frac{F \dot{H}}{\dot{F}}-\frac{F}{3 \dot{F}} \frac{k^{2}}{a^{2}}\right) \Phi_{\mathrm{MG}}-\frac{\kappa^{2} \bar{\rho}}{3 \dot{F}} \delta_{\mathrm{MG}} =0  \nonumber
\end{align}
where
$\kappa_{\mathrm{eff}}^{2}=\frac{\kappa^{2}}{F}\left(1+4 \frac{k^{2}}{a^{2}} \frac{\partial \overline{\ln F}}{\partial R}\right) /\left(1+3 \frac{k^{2}}{a^{2}} \frac{\partial \overline{\ln F}}{\partial R}\right)$.

Here we see the clear difference in the evolution of scalar perturbations. Hence if we can identify the observable quantities that can be constructed using the scalar perturbations, those quantities can be used to detect the signatures of {MG} theories. Here we provide three such observables.

\noindent \textbf{Weak gravitational lensing:} By looking at the spatial part of the geodesic equation for the perturbed metric in the Newtonian gauge, we see that the quantity $\Phi + \Psi$ determines the variation in the path of the photon propagation, which leads to the weak gravitational lensing. Hence weak lensing data can be used to study the spatial dependence of the metric scalar perturbations $\Phi$ and $\Psi$, which we can use to detect the {MG} signatures.

\noindent \textbf{Integrated Sachs-Wolfe effect:}
The integrated Sachs-Wolfe (ISW) effect is a secondary anisotropy of the cosmic microwave
background (CMB), which arises because of the variation in the cosmic gravitational potential
between local observers and the surface of the last scattering. ISW effect is related
to the rate of change of ($\Phi + \Psi$) w.r.t. conformal time ($\eta$). 
ISW effect provides valuable information about the time evolution of the scalar perturbations, especially in
the late accelerating Universe. Even though its detectability is weaker than weak lensing,
it is a powerful tool to study the underlying cosmology. It can be detected using the cross-correlation between the observational data on CMB and large-scale structures.

Recently, it was shown that the difference in the growth
of the scalar perturbations can be used to distinguish $f(R)$ gravity from GR in a model-independent manner using the weak lensing and integrated Sachs-Wolfe effect~\cite{2019-Johnson.Shankaranarayanan-Phys.Rev.D}.

\subsection{Potential to solve long-standing problems in cosmology and particle physics}

The origin of primordial magnetic fields and the origin of the baryon asymmetry of the Universe are the unresolved issues in modern cosmology and particle physics models. Both require physics beyond the standard model and pose an exciting question--- 
are these processes cosmological or particle physics or both? It seems impossible to generate the observed baryon asymmetry within the standard model of particle physics framework. Since both require physics beyond the standard model, there is a tantalizing possibility that the same physics can solve both problems. Recently, it was shown that non-minimal coupling of the electromagnetic field with the Riemann tensor could potentially explain magnetogenesis and baryogenesis~\cite{2020-Kushwaha.Shanki-PRD,2021-Kushwaha.Shanki-PRD}.

As mentioned earlier, the general effective field theory of gravity coupled to the standard model of particle physics was constructed recently~\cite{2019-Ruhdorfer.etal-JHEP}. The authors systematically showed that the first gravity operators appear at mass dimension $6$ in the series expansion, and these operators only couple to the standard model Bosons. This leads to the following action~\cite{2020-Kushwaha.Shanki-PRD}:
\begin{eqnarray}\label{eq:Model_action}
S &= \frac{1}{2 \kappa^2} \int{d^4x \sqrt{-g} \, R} +  \int d^4x \sqrt{-g} \left[  \frac{1}{2} \partial_{\mu}\phi \partial^{\mu}\phi -  V(\phi) \right] \nonumber\\ 
&-\frac{1}{4} \int d^4x \, \sqrt{-g} \, F_{\mu\nu} F^{\mu\nu} - \frac{\sigma}{M^2} \,\int d^4x \, \sqrt{-g} \, R_{\rho\sigma}\,^{\alpha\beta} F_{\alpha\beta} \, \tilde{F}^{\rho\sigma}
\end{eqnarray}
where $R_{\rho\sigma}\,^{\alpha\beta}$ is the Riemann tensor, $A_{\mu}$ is the four-vector potential of the electromagnetic field, $F_{\mu\nu} = \nabla_{\mu}A_{\nu} - \nabla_{\nu}A_{\mu} $ and $\tilde{F}^{\rho\sigma} = \frac{1}{2} \epsilon^{\mu\nu\rho\sigma}F_{\mu\nu} $ is the dual of $F_{\mu\nu}$. $\epsilon^{\mu\nu\rho\sigma} = \frac{1}{\sqrt{-g}}\, \eta^{\mu\nu\rho\sigma}$ is fully antisymmetric tensor, $\eta^{\mu\nu\rho\sigma}$ is Levi-Civita symbol whose values are $\pm1$ and we set $\eta^{0123} = 1 = - \eta_{0123}$. 

Note that in Eq. (\ref{eq:Model_action}), the first three terms correspond to Einstein-Hilbert action, scalar field action, and standard electrodynamics, respectively. However, the presence of the Riemann tensor term breaks the conformal (and parity) invariance of the action. The scalar field 
($\phi$) drives the inflation in the early Universe. 
$M$ is the energy scale, which sets the scale for conformal invariance breaking. We assume that
$10^{-3}  \leq (H_{\rm Inf}/M) \leq 1$ where $H_{\rm Inf} \sim 10^{14}~{\rm GeV}$ is the Hubble scale during inflation. Note that the Riemann coupling is tiny in the current epoch and will significantly contribute only in the early Universe.  

Since the non-minimal coupling term also breaks parity, during inflation, one can generate an appreciable amount of helical fields that seeds large-scale magnetic fields~\cite{2020-Kushwaha.Shanki-PRD}. Interestingly, the generation of the non-zero primordial helical magnetic fields leads to a chiral anomaly, which {results in the imbalance} between left and right-handed fermions. In the presence of an electromagnetic field in curved space-time, the chiral anomaly is given by the following equation~\cite{2021-Kushwaha.Shanki-PRD}:
\begin{align}\label{eq:chiralAnomaly}
 \nabla_{\mu}J_A^{\mu}  = -\frac{1}{384 \pi^2} \epsilon^{\mu\nu\rho\sigma}  R_{\mu\nu\alpha\beta} R^{\alpha\beta}\,_{\rho\sigma} + \frac{e^2}{16 \pi^2} \epsilon^{\mu\nu\alpha\beta} F_{\mu\nu} F_{\alpha\beta}
\end{align}
where $J^{\mu}_A$ is the chiral current. In the case of FRW Universe \eqref{eq:FRW}, the non-zero contribution of the first term arises only in the second-order, like in the case of {CS} gravity~\cite{2006-Alexander.etal-PRL}. However, due to the presence of the magnetic fields, the second term in the RHS of Eq.(\ref{eq:chiralAnomaly}) is non-zero. Thus, the net baryon asymmetry is~\cite{2021-Kushwaha.Shanki-PRD}:
\begin{equation}
\label{eq:n_B-n_CS-definition}
n_B = a(\eta) \langle 0 | J^0_A | 0 \rangle
= \frac{e^2}{4\pi^2} a(\eta) n_{CS}.
\end{equation}
where 
\begin{align}\label{eq:n_cs-relation}
n_{CS} = \frac{1}{a^4} \epsilon_{i j k} \langle 0 | A_i \, \partial_j A_k | 0 \rangle = \frac{1}{a^4}\int_{\mu}^{\Lambda} \frac{dk}{k} \frac{k^4}{2\pi^2} \left(  | A_+ |^2 - |A_-|^2  \right) \, ,
\end{align}
Thus, $n_{CS}$ is non-zero if the primordial magnetic fields are helical, i. e. $|A_+ | \neq |A_-|$, {hence} leading to baryogenesis. Focusing on the modes that leave the horizon around 5 to 10 e-foldings leads to 
the observed amount of baryon asymmetry $\eta_B \sim 10^{-10}$ with the reheating temperature in the range 
$10^{12} - 10^{14}$~GeV~\cite{2021-Kushwaha.Shanki-PRD}.

\section{Future Outlook}

Technological advancements have fueled research to answer some of the fundamental questions of the Universe. GR was proven via the direct detection of GWs from the mergers of the binary BHs and binary neutron stars by the Advanced LIGO and Advanced Virgo detectors. These detections confirmed the prediction of GR and provided the first direct evidence of the existence of stellar-mass BHs. However, the occurrence of singularities at the centers of BHs suggests that GR is inapplicable because of the breakdown of the equivalence principle at the singularities. The fact that these singularities exist indicates that GR cannot be a universal theory of space-time. Many {MG} models are currently proposed, like $f(R)$, Stelle and {CS} gravity. Since the equations of motions of these theories are non-linear and contain {higher derivatives}, we have not obtained spherically symmetric solutions that are distinct and different from GR. Like for instance, {Schwarzschild is also a solution for $f(R)$, Stelle and {CS} gravity}. The current challenge is to obtain exact black hole solutions that are unique to modified theories. Non-trivial solutions are obtained in the literature only for a specific form of $f(R)$~\cite{2020-Xavier.etal-CQG}. 

In the low-energy limit, the theoretical and observational challenges faced by the $\Lambda$CDM model also indicate that we might have to look beyond GR as the underlying theory of gravity. Here again, the challenge is to obtain decelerated to accelerated expanding Universe for generic {MG} theories without instabilities. 

Testing {GR} using GW observations relies on combining template matching and polarization measurements. Currently, template matching is used to match the GW signals with numerical relativity waveforms to constrain deviations from GR. However, as the number of detectors increases and their sensitivity to the ring-down region, alternative methods are needed to test for deviations in GW signals. One must obtain a finite number of parameters vanishing for GR and non-vanishing for {MG} theories. In Refs.~\cite{2019-Bhattacharyya.Shanki-PRD,2019-Shankaranarayanan-IJMPD} one such parameter was suggested. More such parameters are required in the strong and weak gravity limits that can be tested in future missions like Cosmic Explorer,  Einstein telescope, E-LISA, Euclid, GAIA, LOFAR, SKA, uGMRT. 

GWs carry a lot of energy, much more than radio waves. For instance, the GWs that we receive on Earth have the energy of the order of {$10^{20}~{\rm Jansky}$}~\cite{2009-Sathyaprakash.Schutz-LRR}. Recently, it was shown that the energy carried by GWs could have observable effects like in Fast Radio Bursts~\cite{2022-Kushwaha.etal}. Such high-energy astronomical phenomena are a rich source of indirect signatures for {MG} theories.

{Two important topics we have not discussed in this review are black hole thermodynamics in MG theories and infrared corrections combining quantum mechanics and gravity. Hawking established the connection between black-hole mechanics and ordinary
thermodynamics by showing that the surface
gravity of the event horizon equals $2\pi$ times the standard
temperature of the radiation emitted by the black hole to infinity
\cite{1975-Hawking-CMP}. The original derivation of the laws of
black hole thermodynamics has nothing in common with statistical
mechanics. Still, there is a general belief that a connection
exists at the quantum level. In the leading order,
several approaches using completely different microscopic degrees of
freedom lead to Bekenstein-Hawking entropy. However, other approaches lead to disparate quantum corrections to Bekenstein-Hawking entropy~\cite{2001-Wald-LivingReviewsinRelativity,2005PageNewJournalofPhysics,2007CarlipJournalofPhysicsConferenceSeries,Carlip2009,Das2010}. Thus, it is unclear whether all the laws of black-hole mechanics will be satisfied for MG models. For recent reviews, see~\cite{2018-Wall,2019-Sarkar-Gen.Rel.Grav.}} 

{Quantum mechanics requires the well-defined notion of generators of translation. While this is in-general true for well-defined space-times that are asymptotically flat, it is \emph{not guaranteed} for asymptotically non-flat space-times~\cite{kempf1996noncommutative}. This has led to conjectures that the commutation relation of a quantum particle will be modified in curved geometry~\cite{petruzziello2021gravitationally,wagner2022relativistic}. Recently, 
it has been demonstrated that the infrared modifications to the position-momentum algebra are proportional to the curvature invariants (like Ricci scalar, Kretschmann scalar)~\cite{2022-Gattu-Shanki}. This may provide a route to understanding whether the GR asymptotic symmetry group should be larger or smaller than the original Bondi–Metzner–Sachs (BMS) group~\cite{Barnich:2009se}.}

\begin{acknowledgements}
The authors fondly remember Prof. Thanu Padmanabhan, who always encouraged his students \emph{to create their own paths}. The authors thank S. M. Chandran, A. Kushwaha, and Urjit Yajnik for their comments on the earlier draft. JPJ is supported by the CSIR fellowship. This work is supported by SERB-MATRICS and ISRO Respond grants. 
\end{acknowledgements}

\section*{Data availability}
Data sharing not applicable to this article as no datasets were generated or analysed during the current study.

\input{references.bbl}
\end{document}

%% file: references.bbl
%